\documentclass[a4paper,fleqn,usenatbib]{mnras}

\usepackage{newtxtext,newtxmath}
 
\usepackage[T1]{fontenc}
\usepackage{ae,aecompl}

\usepackage{graphicx}	
\usepackage{amsmath}	
\usepackage{amssymb}	
\usepackage{pdflscape}

\newcommand{\msun}{M$_{\odot}$} 
\newcommand{\kms}{\,km\,s$^{-1}$} 
\newcommand{\ha}{H$\alpha$} 

\title[\ha~correlation with inclination and a low-mass BH]{A correlation between \ha~trough depth and inclination 
in quiescent X-ray transients: evidence for a low-mass black hole in GRO J0422+32}
\author[J. Casares et al.]{J. Casares$^{1,2}$\thanks{E-mail: jorge.casares@iac.es},
T. Mu\~noz-Darias$^{1,2}$,  M.A.P. Torres$^{1,2}$, D. Mata S\'anchez$^{1,2}$,  C.T. Britt$^{3}$,  
\newauthor M. Armas Padilla$^{1,2}$, A. \'Alvarez-Hern\'andez$^{1,2}$, V.A. C\'uneo$^{1,2}$, J.I. Gonz\'alez Hern\'andez$^{1,2}$,
\newauthor F. Jim\'enez-Ibarra$^{4}$, P.G. Jonker$^{5,6}$, G. Panizo-Espinar$^{1,2}$, J. S\'anchez-Sierras$^{1,2}$ and 
\newauthor I.V. Yanes-Rizo$^{1,2}$  
\\
$^{1}$Instituto de Astrof\'isica de Canarias, E-38205 La Laguna, Tenerife, Spain\\
$^{2}$Departamento de Astrof\'isica, Universidad de La Laguna, E-38206 La Laguna, Tenerife, Spain\\
$^{3}$Space Telescope Science Institute, 3700 San Martin Dr, Baltimore, MD 21218, USA\\
$^{4}$Australian Astronomical Optics, Macquarie University, 105 Delhi Rd, North Ryde, NSW 2113, Australia\\
$^5$Department of Astrophysics/IMAPP, Radboud University, PO~Box 9010, NL-6500 GL, Nijmegen, the Netherlands\\
$^6$SRON, Netherlands Institute for Space Research, Niels Bohrweg 4, NL-2333 CA Leiden, the Netherlands
}
\date{Accepted XXX. Received YYY; in original form ZZZ}

\pubyear{2022}

\begin{document}
\label{firstpage}
\pagerange{\pageref{firstpage}--\pageref{lastpage}}
\maketitle

\begin{abstract}
We present a new method to derive binary inclinations in quiescent black hole (BH) X-ray transients (XRTs), based 
on the depth of the trough ($T$) from double-peaked \ha~emisison profiles arising in accretion discs. We find that 
the inclination angle ($i$) is linearly correlated with $T$ in phase-averaged spectra with sufficient orbital coverage 
($\gtrsim$50 per cent) and spectral resolution, following $i ({\rm deg})=93.5\times T +23.7$.
The correlation is caused by a combination of line opacity and local  broadening, 
where a leading (excess broadening) component scales with the deprojected velocity of the outer disc. 
Interestingly, such scaling allows to estimate the fundamental ratio $M_{1}/P_{\rm orb}$ by simply resolving 
the intrinsic width of the double-peak profile.  We apply the $T-i$ correlation to derive binary inclinations for 
GRO J0422+32 and Swift J1357-0933, two BH XRTs where strong flickering activity has hindered determining their 
values through ellipsoidal fits to photometric light curves.  
Remarkably, the inclination derived for GRO J0422+32 ($i=55.6\pm4.1^{\circ}$) implies a BH mass of 
$2.7^{+0.7}_{-0.5}$ \msun~thus placing it within the gap that separates BHs from neutron stars. 
This result  proves that low-mass BHs exist in nature and strongly suggests that the so-called "mass gap" is mainly 
produced by  low number statistics and possibly observational biases. On the other hand, we find that  Swift J1357-0933 
contains a $10.9^{+1.7}_{-1.6}$  \msun~BH, seen nearly edge on ($i=87.4^{+2.6}_{-5.6} $ deg). Such extreme inclination, 
however, should be treated with caution since it relies on extrapolating the 
$T-i$ correlation beyond $i\gtrsim75^{\circ}$, where it has not yet been tested.
\end{abstract}




\section{Introduction}
\label{intro}

Galactic black holes (BHs) compose a unique benchmark to investigate BH formation via stellar evolution. Many 
are found in X-ray transients (XRTs), a sub-class of X-ray binaries that show episodic X-ray outbursts, triggered by 
mass accretion (e.g. \citealt{mcclintock06, belloni11}). 
Weighing BHs in XRTs hinges on detecting the dim companion star at optical and/or near-infrared (NIR) wavelengths 
during quiescence,  i.e. when accretion is significantly reduced and the X-ray luminosity is low. 
Dynamical information (orbital period $P_{\rm orb}$ and radial velocity semi-amplitude $K_2$) is 
extracted from the radial velocity curve of the companion, while modeling the ellipsoidal light curve (shaped by the tidally 
distorted surface) allows constraining the orbital inclination $i$. $P_{\rm orb}$ and $K_2$ are usually determined with exquisite 
precision but the measurement of $i$ is more prone to be affected by systematics. In particular, the presence of aperiodic 
variability from the accretion disc (i.e. flickering; \citealt{zurita03,cantrell10}) veils and distorts the ellipsoidal light curve, 
leading to biased inclination determinations. This can have a critical impact on BH mass calculations because of the cubic 
dependence of the mass function on $\sin i$ (e.g. \citealt{kreidberg12}). We refer the reader to \cite{casares-jonker14} for 
a review on BH mass measurements and the possible systematics involved. 

There are currently 68 XRTs hosting potential BHs, but only 19 have been successfully confirmed through dynamical studies 
(see the online version of the BlackCAT catalogue\footnote{http://www.astro.puc.cl/BlackCAT/index.php}; \citealt{corral16}).  
In an effort to expand the BH census further, new scaling relations and survey strategies based on the strong disc 
\ha~emission line have been proposed \citep{casares15, casares16, casares18, casares-torres18}. 
Building upon these relations, indirect evidence for dynamical BHs has been reported in four additional XRTs 
whose faint quiescent counterparts ($r\gtrsim22$) have so far hampered the detection of stellar absorption features. 
These are Swift J1357-0933 \citep{mata15,casares16}, 
KY TrA \citep{zurita15}, Swift J1753.5-127 \citep{shaw16} and MAXI J1659-112 \citep{torres21}. 

Here in this paper we investigate a new scaling relation between the depth of the \ha~line trough ($T$) and binary inclination. 
Such scaling makes it possible to circumvent flickering biases, providing inclination measurements for XRTs with highly veiled 
or undetected companions. This is because flickering contaminates equally the entire double-peak profile thus cancelling out 
possible variations imparted on the depth of the line core \citep{hynes02}.
In Sections \ref{sec:signpost} and \ref{sec:calibration} we introduce our \ha~fitting model and reference sample, 
respectively,  while Section \ref{sec:correlation} presents the derived $T-i$ correlation. In Section \ref{sec:cvs} 
we compare our results on $T$ and $i$ obtained using XRTs with those from cataclysmic variables. 
The origin of the $T-i$ correlation is investigated in Section \ref{sec:modelling} using optically thick 
line profile simulations, while it is applied in Section \ref{sec:2_bhs} to two XRTs with poor existing constraints. 
In Section \ref{sec:discussion} we present the BH masses derived from the new inclination 
measurements and discuss their implications. Notably, we find that one XRT hosts a BH sitting in the 
mass gap with high confidence. Finally, we summarize our results in Section \ref{sec:conclusions}. 

\section{A signpost of binary inclination}
\label{sec:signpost}

It has long been recognized that the central depression in double-peaked accretion disc lines deepens with inclination 
(see for example \citealt{horne-marsh86}), but no attempt has been made yet to exploit this feature to derive binary inclinations. 
We propose to do so here by fitting a simple analytical model to \ha~profiles from a sample of quiescent XRTs
with reliable inclinations. The observed profiles can be described by a symmetric model consisting of 
two Gaussians of equal height $h$ and standard deviation $\sigma$. The line flux is, thus, given by 

\begin{equation}
f(x)= h~\left[e^{-\left(\frac{x-x_{0}-DP/2}{\sqrt{2}\sigma}\right)^{2}} + e^{-\left(\frac{x-x_{0}+DP/2}{\sqrt{2}\sigma}\right)^{2}}\right] ,
 \label{eq:model}
\end{equation}

\noindent
where  $x$ is the velocity, $x_{0}$ the velocity displacement of the model centroid 
relative to the \ha~rest wavelength (6562.76 \AA) and $DP$ the double peak separation. 
The flux at the central depression or trough ($f_{T}$) is set 
by $f(x=x_{0})$, i.e. 
 
\begin{equation}
f_{T}=2~h~e^{-\left(\frac{DP}{2\sqrt{2}\sigma}\right)^{2}}.
\label{eq:depression}
\end{equation}

\noindent
The depth of the central depression is given by $h-f_{T}$  and, since the full-width at half-maximum of each Gaussian 
is $W=2\sqrt{2 \ln 2}~\sigma$, we can express the dimensionless depth of the trough (i.e. normalized to the double 
peak height) as 

\begin{equation}
T=\frac{h-f_{T}}{h}=1-2^{1-\left(\frac{DP}{W}\right)^{2}}
\label{eq:trough}
\end{equation}

\noindent
Eq.~\ref{eq:trough} provides a 
way to parametrize the depth of the central line depression by fitting a simple analytic model to the data. In the next 
two sections we will apply this relation to extract $T$ values from a sample of BH XRTs with secure inclinations, 
what we name the  {\it calibration sample}.  However, before proceeding, it is important to consider two potential 
sources of  systematics that may affect the $T$ measurements, namely the impact of limited instrumental resolution 
and orbital coverage. These effects are thoroughly investigated in  Appendix~\ref{ap:resolution} and 
Appendix~\ref{ap:orbital} while here we simply outline the main results. These are: (1) $T$ values can be reliably 
measured if the observed profiles are fitted with 2-Gaussian models (i.e. eq.~\ref{eq:model}) that have been previously 
degraded to the instrumental resolution. We find that the measurements thus obtained are not significantly biased 
if instrumental resolution ($\Delta V_{\rm res}$) remains better than $\approx0.5 \sqrt{(DP^{2}-W^{2})}$ or, in other words, 
the here called {\it scaled-resolution} parameter $\Delta \equiv \Delta V_{\rm res}/\sqrt{(DP^{2}-W^{2})}$ is $ \lesssim 0.5$.  
(2) $T$ describes a double-humped modulation with orbital phase, reminiscent of ellipsoidal light curves. 
This is an ubiquitous feature of quiescent XRTs and is explained  by the periodic motion 
of S-waves (tied to the hot-spot and the companion star) across the line profile. 
As a consequence, $T$ values obtained from spectral averages with limited phase coverage could, in principle, be 
biased.  To mitigate this effect we will hereafter focus on $T$ measurements obtained from data with 
sufficiently large ($\gtrsim50$ per cent) orbital coverage and, otherwise, warn about the impact of possible systematics.
  
 \section{The calibration sample}
\label{sec:calibration}

In an effort to provide a list of accurate inclinations we have selected six BH XRTs with inclination measurements based on 
ellipsoidal light curve modeling during passive state periods (whenever available) and corrected for accretion disc 
contamination, either through multi-band photometric fits or simultaneous/contemporaneous spectroscopy. 
We refer the reader to  Appendix~\ref{ap:dismissed_inclination} for our assessment of other inclination values reported in 
literature that are not considered here. The list of selected targets with our favored inclinations is: 

\begin{itemize}
\item{}{{\bf A0620-00 (=V616 Mon; hereafter A0620):}} \cite{cantrell10} present ellipsoidal fits to $VIH$ light curves selected 
         during passive state (i.e. with minimum flickering), corrected for disc contamination through simultaneous optical and 
         NIR spectroscopy.They obtain $i=51.0\pm0.9^{\circ}$.  In a subsequent study, \cite{vangrunsven17} perform fits to 
         the same light curves but using different modeling and fitting strategy, resulting in $i=54.1\pm1.1^{\circ}$. The 
         difference between the two values likely reflects the impact of systematics associated with different modeling strategies. 
         Since the measurements are not independent, we adopt the unweighted mean of the two values and a conservative 
         uncertainty that encompasses them i.e. $i=52.6\pm2.5^{\circ}$. For the line profile fit we use 78 spectra obtained with 
         the Gran Telescopio Canarias (GTC)  on the nights of 2012 Dec 5-6 and 2013 Jan 7 at 140 \kms~spectral 
         resolution\footnote{Here and in what follows, we quote instrumental resolution values measured from the $FWHM$ of 
         sky or arc lines obtained on the same night and with identical slit width and instrument configuration as the data. Only 
         in the case of digitized spectra and those  provided by others we adopt instrumental resolution values quoted in the relevant 
         papers where these spectra are presented.} (Gonz\'alez Hern\'andez et al., in preparation). 
         These spectra provide full coverage of the binary orbit. 
         To check the consistency of trough depth measurements across different epochs and data sets we have also included 
         \ha~observations reported in \cite{marsh94}. These consist of medium resolution spectra (70 \kms) obtained with the 
         William Herschel Telescope (WHT) on the nights of 1991 Dec 31 and 1992 Jan 1 and covering an entire binary orbit. 
         Unfortunately,  the original data are not available from the Issac Newton Group Archive and we had to digitize the 
         averaged \ha~profile from fig 12 in \cite{marsh94}. Likewise, we have digitized the average \ha~profile shown in fig. 2 
         of \cite{neilsen08} because the original data are not available  from the National Optical Astronomy Observatory 
         (NOAO) Service Archive. This spectrum results from observations obtained with the Clay Magellan telescope on the 
         nights of 2006 April 14-16  at 130 \kms~resolution. 
         
\item{}{{\bf GRS 1124-684 (= N Mus 91):}} We choose  the inclination value ($i=43.2^{+2.1}_{-2.7}$ deg) reported in a 
          detailed study by \cite{wu16}. This is based on ellipsoidal fits of optical-NIR light curves, 
        selected in passive state over 24 years and corrected for accretion disc contamination (as derived from simultaneous 
        spectroscopy during the 2009 database).  A total of 31 \ha~spectra from \cite{casares97} were employed for the 
        Gaussian model fit.  These spectra were collected with the New Technology Telescope (NTT) during 1993-1995 at 
        90 \kms~resolution and cover a complete orbital cycle. To check for possible changes in trough depth throughout different 
        epochs we have also included 40 \ha~spectra obtained in 2009 with the Clay Magellan Telescope at 46 \kms~resolution  
        \citep{wu15}, and 17 Very Large Telescope (VLT) spectra from 2013 at 43 \kms~resolution \citep{gonzalez17}. Each of 
        these data sets covers $\approx$70-75 per cent of the binary orbit.

\item{}{{\bf GS 2000+25 (=QZ Vul; hereafter GS2000):}} As for A0620 we quote the unweighted mean of the inclination values 
          reported by \cite{callanan96b} and \cite{ioannou04}. Although only {\it pure} ellipsoidal models were fit to multi-wavelength 
          light curves, both works discuss the impact of disc light contamination in the results. In the case of \cite{callanan96b} they 
           find this to be negligible after extrapolating the optical (spectroscopic) veiling to 
           their $J-$ and $K$-band light curves and obtain $i=65\pm9^{\circ}$. Regarding \cite{ioannou04}, we adopt  
           a uniform inclination distribution limited by their non-eclipse and no-disc models, i.e. $i=59-81^{\circ}$.  
          We therefore take $i=67.5\pm5.7^{\circ}$ as our best inclination for GS2000 
           where the error has been derived by randomizing the normal and flat inclination distributions from \cite{callanan96b} and 
           \cite{ioannou04}, respectively. The line profile fit was performed over an orbital average Keck spectrum from 
           \cite{filippenko95}, with 120 \kms~resolution and full phase coverage. 
           Despite their lower quality, we also include 27 WHT spectra from \cite{casares95a} for an independent 
           determination of the double peak trough. These have 196 \kms~resolution and cover the entire binary orbit. 
 
\item{}{{\bf XTE J1118+480 (=KV UMa; hereafter J1118):}} \cite{gelino06} fit (non-simultaneous) 
         $BVRJHK$ light curves while accounting for disc contamination and find $i=68\pm2^{\circ}$. \cite{khargharia13}, 
         on the other hand, fit $H$-band light curves, with disc contamination estimated from contemporaneous NIR spectroscopy, 
         and obtain $i=68-79^{\circ}$. More recently, \cite{cherepashchuk19} fit optical and simultaneous NIR light curves, allowing for 
         accretion disc contribution, and find $i=74\pm4^{\circ}$. We adopt the unweighted mean and standard deviation after 
         randomizing these three independent measurements, assuming  Gaussian distributions for \cite{gelino06} and 
         \cite{cherepashchuk19}, and a flat distribution for \cite{khargharia13}. Our favoured inclination is, therefore, $i=71.8\pm1.8^{\circ}$. 
         A total of 162 GTC spectra of J1118 \citep{gonzalez12,gonzalez14} 
         were employed for the line profile fit. The spectra were obtained on the nights of 2011 Feb 7 \& 8, 2011 Apr 25 and 2012 Jan 12 
         and they all possess an instrumental resolution of 120 \kms . Because each night covers a full orbital cycle these were 
         treated as four independent epochs.  We also include the average of 72 Keck spectra collected in 2004 at 50 
         \kms~resolution \citep{gonzalez06, gonzalez08}. The orbital coverage of the Keck epoch is also complete.

\item{}{{\bf XTE J1550-564 (=V381 Nor; hereafter J1550):}}  \cite{orosz11} present a refined dynamical study with improved 
          ellipsoidal fits to $JK_{\rm S}$ light curves, accounting for disc contamination, and find $i=74.7\pm3.8^{\circ}$. We take 
          this inclination as the best determination available for this system. For the \ha~fit we use the average of 16 Magellan spectra 
          from \cite{orosz11} with 55 \kms~resolution. The spectra only provide a limited $\sim$30 per cent orbital coverage, centered 
          at phases 0 and 0.7.    
           
\item{}{{\bf MAXI J1305-704 (hereafter J1305):}} We adopt the result of \cite{mata21} based on ellipsoidal fits to $gri$ light curves, 
         including the  contribution of an accretion disc. They report  $i=72^{+5}_{-8}$ deg. We use the average of 16 VLT 
         spectra with 140 \kms~resolution and complete phase coverage \citep{mata21} for the line profile analysis.   

\end{itemize}

In order to extend our analysis to low inclinations we also consider 20 high-resolution (7 \kms) VLT spectra of 
 the neutron star XRT Cen X-4, reported in \cite{casares07}. These spectra were obtained around quadrature orbital phases 
 $\sim$0.25 and  $\sim$0.75. Three accurate inclination values are reported in literature for Cen X-4. \cite{khargharia10} find 
 $i=35^{+4}_{-1}$ deg after fitting ellipsoidal models to an $H$-band light curve, with veiling correction derived through 
 non-simultaneous NIR spectroscopy. Meanwhile, \cite{hammerstein18} obtain $i=34.9^{+4.9}_{-3.6}$ deg by modeling $JHK$ 
 light curves, allowing for an accretion disc contribution. Finally, following a completely different approach, \cite{shahbaz14} report  
 $i=32^{+8}_{-2}$ deg by modeling the absorption line profiles of the companion star using Roche tomography techniques. 
 As in the case of the BH calibrators we take the unweighted average and error through randomizing the three independent values, 
 i.e.  $i=34.0\pm2.6^{\circ}$. 
 
 \section{A correlation between \ha~trough and inclination}
\label{sec:correlation}

\begin{figure}
	\includegraphics[angle=-90,width=\columnwidth]{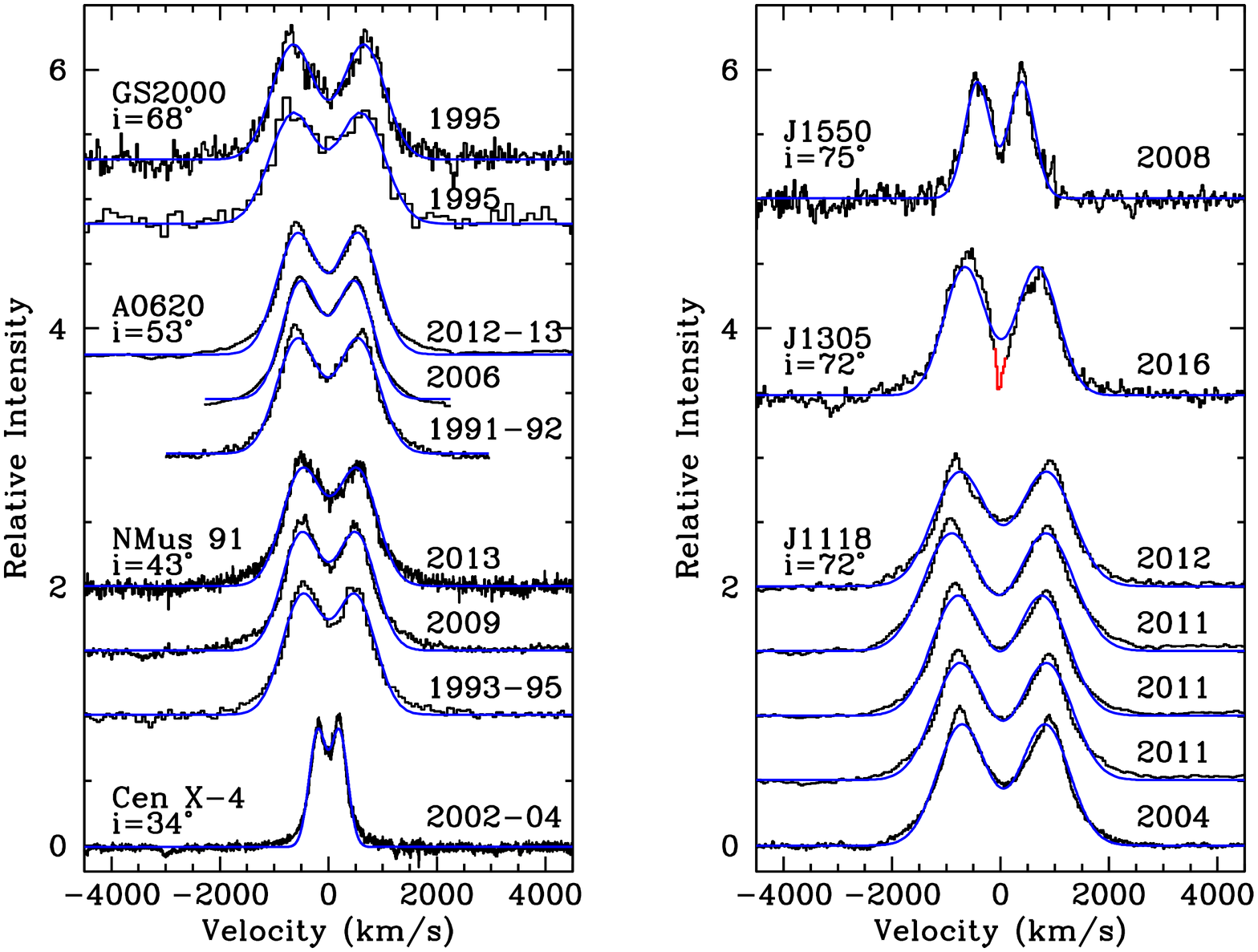}
    \caption{
    Orbital average \ha~spectra of our seven calibrators and their best 2-Gaussian model fits (blue). Different epochs/databases are shown for NMus 91, A0620, GS2000 and J1118.  Binary inclinations are indicated for each source. 
 In the case of the J1305 profile we mark in red the (masked) region contaminated by an interloper star (see Mata S\'anchez et al. 2021).
 }
    \label{fig:fig1}
\end{figure}

We fit our symmetric 2-Gaussian model to the orbital averaged \ha~ profiles of the seven calibration sources and apply 
eq.~\ref{eq:trough} to derive $T$, the depth of the line trough (see Fig.~\ref{fig:fig1}). As referred in Section~\ref{sec:signpost}, 
the models were always degraded to the instrumental resolution of each spectrum through convolution with a Gaussian  
with full-width at half-maximum $FWHM=\Delta V_{\rm res}$.  Different fits were performed for each epoch independently.  
The spectra have the continuum level subtracted and their peak intensities normalized to unity.  The fits have 
been performed in a window of $\pm$10000 km s$^{-1}$ centered on \ha~after masking the neighboring HeI line at 
$\lambda$6678. We adopt 1-$\sigma$ formal errors on the fitted parameter as derived through $\chi^2$ minimization. 

In the case of the J1305 spectrum we followed  \cite{mata21} and mask part of the line core to 
remove the contamination produced by a well documented interloper star. 
In order to establish the optimal mask size we studied the impact of different widths on $T$ values, measured from a 
simulated profile with interloper contamination.This was constructed by adding a Gaussian absorption (with 
$FWHM= \Delta V_{\rm res}$ and centered at rest velocity) to the core of the J1118 \ha~line. 
The depth of the contaminating \ha~absorption is scaled to the interloper's contribution as in \cite{mata21}. 
The choice of J1118 is motivated by the fact that both its spectral resolution and trough depth are nearly 
identical to those in J1305. By performing this test we find that the $T$ values obtained from the simulated 
profile are always larger than in the original J1118 spectrum, irrespectively of the size of the mask. 
The closest match is obtained for a mask width of 1.5$\times$$\Delta$$V_{\rm res}$, resulting in $T$ being overestimated by 
0.012. We, therefore, decided to adopt a width mask of 1.5$\times$$\Delta$$V_{\rm res}$ in our fit to the J1305 profile 
and applied a small shift correction of  -0.012 to the resulting $T$ value. We also increased the formal $T$ uncertainty by 
adding quadratically 0.012  in order to account for possible systematics introduced by the mask. 

Regarding J1550, we note that the average spectrum is affected by limited phase coverage and thus, 
as shown in Appendix~\ref{ap:orbital}, the $T$ measurement could be biased. However, the 16 individual spectra 
are evenly distributed around phases 0 and 0.7 (i.e. the minimum and maximum of the $T$ orbital modulation) 
 and, therefore, we expect our $T$ determination not to differ much from its orbital average. 
Nonetheless, to account for possible systematics we have artificially increased the formal $T$ uncertainty (0.012) 
by adding quadratically 0.038 or 25 per cent of the amplitude expected from the orbital modulation.

The analysis of the Cen X-4 profile deserves further consideration. Cen X-4 is renowned for the presence of a very strong 
\ha~emission component, associated to the irradiated companion star \citep{torres02, davanzo05, davanzo06}. 
To minimize the impact of such component  in the measurement of $T$ we produced a composite profile  
by merging two spectral halves with positive (red) and negative (blue) velocities. The blue part of the profile is obtained after 
co-adding VLT spectra taken at phases 0.1-0.4 (when the companion S-wave is located on the red peak)  while the red profile 
is obtained from spectra within phases 0.6-0.9 (when it is placed on the blue peak). The 2-Gaussian model fit to the composite 
profile yields $T=0.147\pm0.001$, but we realize this is an upper limit to the mean orbital value since our spectra sample phases 
close to the $T$ maxima (see Appendix~\ref{ap:orbital}). 
A $T$ correction can be worked out by simulating a double sine-wave modulation of amplitude $\Delta T=\pm0.15$ and maxima 
at phases 0.2 and 0.7. We find that the $T$ value obtained by sampling phases 0.1-0.4 and 0.6-0.9 results in an overestimate 
of the phase average by +0.06. On this ground, we correct our previous determination by -0.06 and adopt a conservative 
errorbar that encompasses it, i.e.  $T=0.09 \pm 0.06$.

\begin{figure}
	\includegraphics[angle=-90,width=\columnwidth]{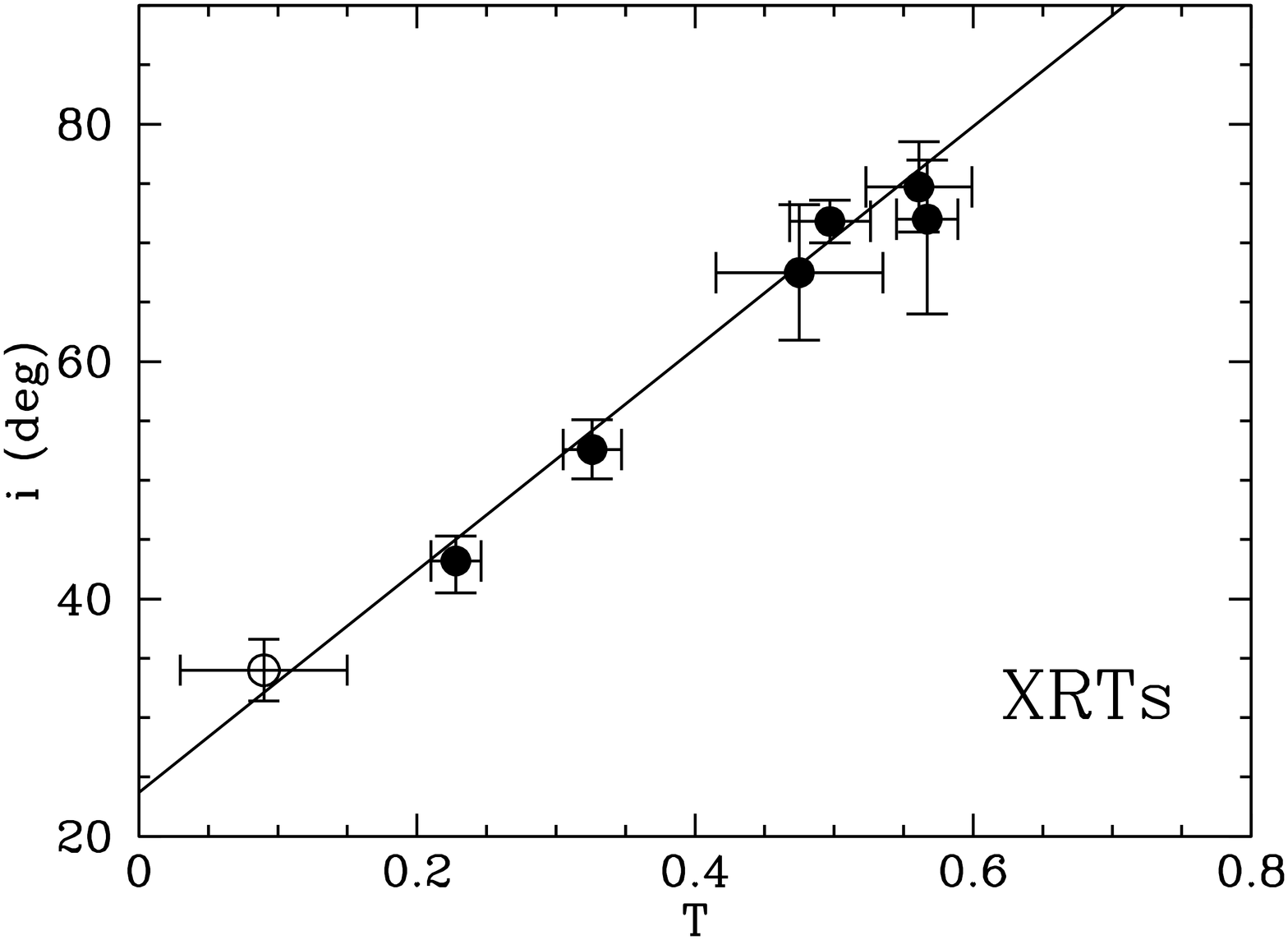}
    \caption{
    Empirical correlation between the depth of the double peak trough ($T$) and binary inclination ($i$) for our seven calibrators. 
    The open circle marks the position of Cen X-4. 
    }
    \label{fig:fig2}
\end{figure}

The final list of $T$ values and inclinations for the seven calibration sources is given in Table~\ref{tab:tab1} and plotted in 
Fig.~\ref{fig:fig2}.  Note that Table~\ref{tab:tab1} also provides the weighted mean of $T$ values (highlighted in bold face) for 
systems with  multiple epochs. We regard these as the best possible determinations of trough depth since  
 the effect of episodic variability is averaged out.  A0620, N Mus 91 and J1118 contain the largest number of epochs (3-5), 
 spanning over 1-2 decades, and henceforth we take their  standard deviations as representative of  the intrinsic variability 
 in $T$ measurements of quiescent XRTs, i.e. $\Delta T \approx 0.02-0.03$.  We also emphasize that the 
 {\it scaled-resolution parameter} is $\Delta \lesssim 0.2$ in all cases thus implying that the $T$ values quoted in 
 Table~\ref{tab:tab1} are not affected by instrumental resolution (see  Appendix~\ref{ap:resolution}). 

The points displayed in Fig.~\ref{fig:fig2} draw a clear linear track.  A least-squares linear fit with the newly scaled error bars yields 

\begin{equation}
i (deg) = 93.5(6.5) ~T + 23.7(2.5)
\label{eq:correlation}
\end{equation}

\noindent 
with a Pearson correlation coefficient $r=0.992$. 
In order to estimate the uncertainty in $i$ implied by this relation we have computed the difference with respect to the true 
observed  inclinations for our seven calibrators. The distribution of differences  is well fit by a normal function with 
$\sigma (i)=2^{\circ}$, which reflects the typical inclination uncertainty drawn by the correlation. 
Note here that we are not attempting to model a more physically motivated variation of $T$ with $\sin i$ because 
this follows a rather more complex (unknown) function than the simple linear correlation described by eq.~\ref{eq:correlation}. 
In any case, the calibration sample covers a wide range of inclinations between $i\sim 35-75^{\circ}$ and thus we believe our  
linear correlation is accurate at least within this interval. 

 \section{Comparison with cataclysmic variables}
\label{sec:cvs}

For the sake of comparison,  we have also obtained $T$ values for eight cataclysmic variables (CVs) from a database 
collected in \cite{casares15}. All the CVs are eclipsing and, therefore, possess very precise inclinations through 
light curve modeling of the white dwarf and/or hot-spot eclipses. 
 The sample consists of three WZ Sge stars (WZ Sge itself, SDSS J103533.02+055158.3 - J1035 hereafter - and 
 SDSS J143317.78+101123.3 - J1433 hereafter) which have large outburst amplitudes and decade-long recurrence times 
 similar to XRTs. We also include five other more frequently outbursting dwarf novae: 
  U Gem, HT Cas, OY Car, IP Peg and  CTCV J1300-3052 (J1300 hereafter).  
   For want of a better term, we refer to these five as {\it dwarf novae} as
   distinct from WZ Sge stars. 

\begin{figure}
	\includegraphics[angle=-90,width=\columnwidth]{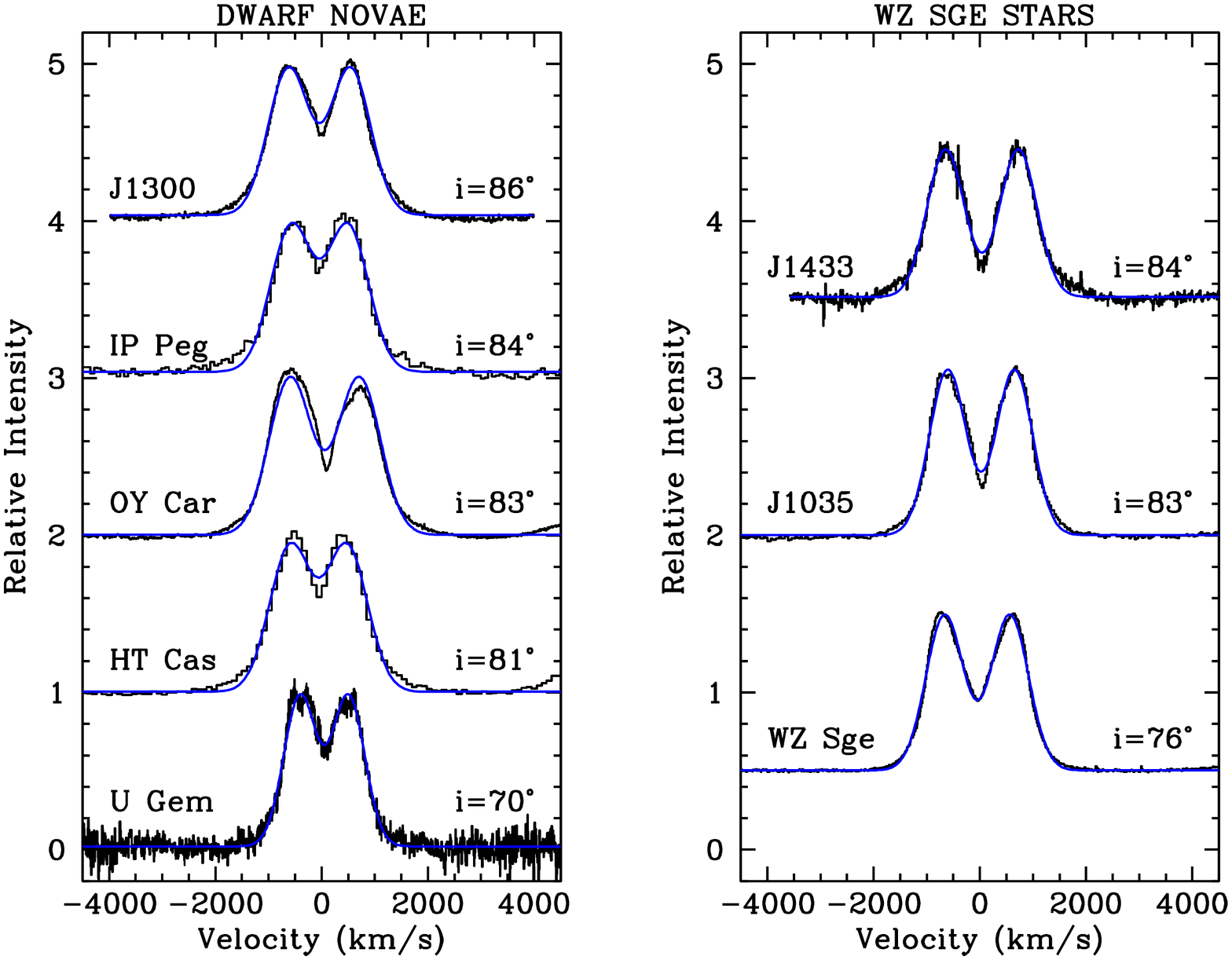}
    \caption{
    Symmetric 2-Gaussian fits to \ha~profiles of a sample of dwarf novae. The five (frequently outbursting) dwarf novae 
    are shown in the left panel, while the right panel displays the three WZ Sge stars.
   Binary inclinations are indicated for each source. 
    }
    \label{fig:fig3}
\end{figure}

\begin{figure}
	\includegraphics[angle=-90,width=\columnwidth]{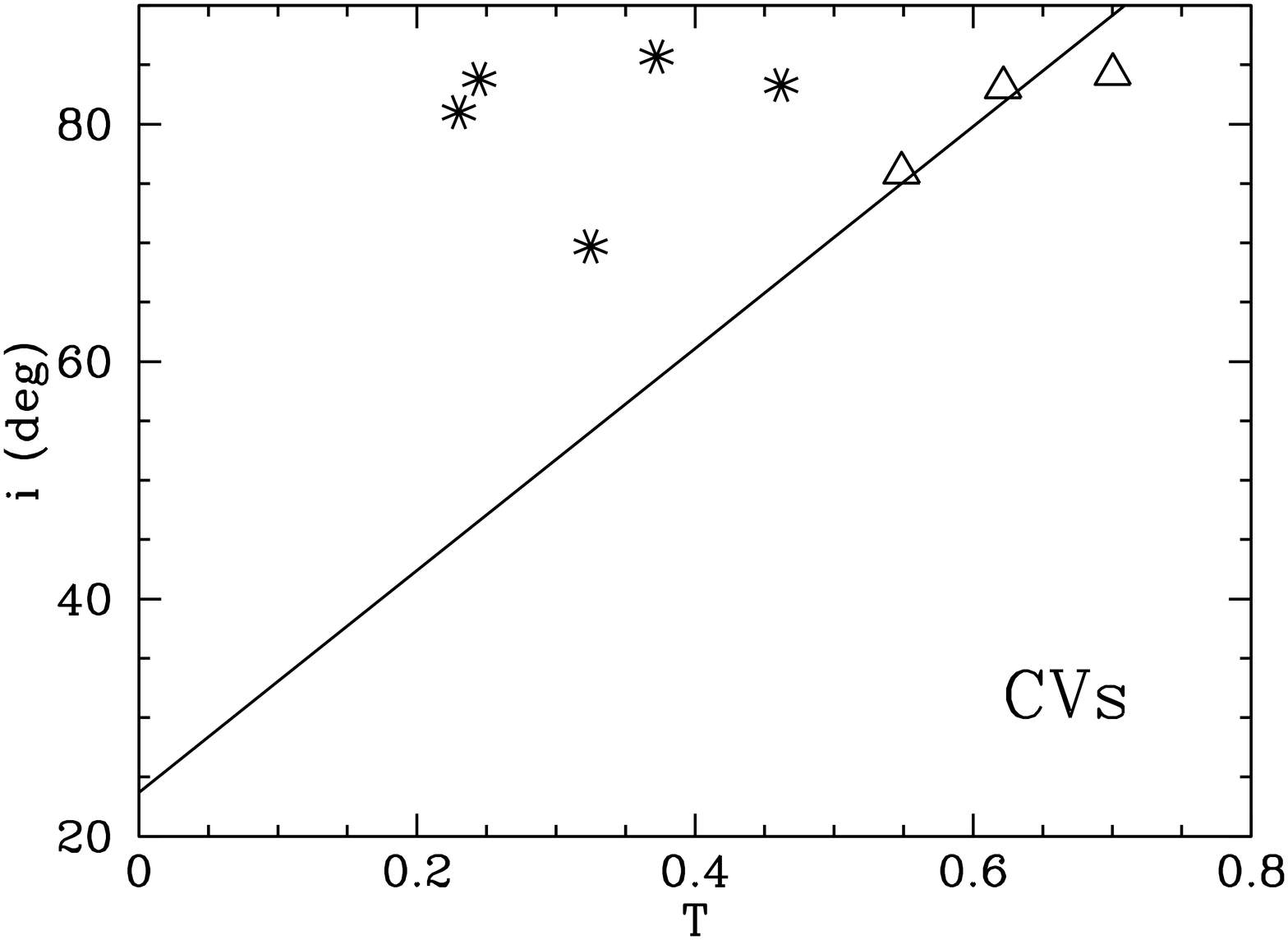}
    \caption{
    Trough depth measured from a sample of CVs, compared to the $T-i$ correlation for XRTs (solid line). 
    Open triangles mark the  position of WZ Sge stars while asterisks those of  the five dwarf novae. 
    }
    \label{fig:fig4}
\end{figure}

For the 2-Gaussian fit analysis we used orbital averages of the spectra reported in Table 2 of \cite{casares15}. 
It should be noted that the \ha~profile in J1035  is embedded in broad absorption wings, caused by superposition 
of the white dwarf spectrum. In this case, we removed the white dwarf contamination  
by subtracting a DA spectrum with $T_{\rm eff}=12000$ K and $\log g=8.0$, scaled to 85 per cent of the total flux (see 
\citealt{southworth06}) and degraded to the spectral resolution of J1035. Table \ref{tab:tab2}  gives details of the CV spectra 
and fitting results, while Fig.~\ref{fig:fig3} presents the data and model fits. A comparison of the measured CV trough 
depths with the $T-i$ correlation for XRTs is displayed in Fig.~\ref{fig:fig4}.

 Interestingly, we observe that trough depths in the five dwarf novae are systematically lower than predicted by the $T-i$ correlation  
 By contrast, WZ Sge-type stars appear  to fit well in the upper side of the correlation. Unfortunately, both the sample size and 
 inclination range are too limited to draw any firm conclusion. Here we simply  
 speculate with the possibility that non WZ Sge-type  dwarf novae have too short outburst recurrence 
 periods ($\sim$tens of days) for accretion discs to settle down in full quiescence. For example, the U Gem database 
 was obtained 18 days after the end of a long outburst while the WZ Sge data $\sim$18 years past the previous 1978 outburst. For 
 comparison, a 20 year monitoring campaign on the BH XRT V404 Cyg indicates that it takes $\approx$4 years for the accretion disc 
 to shrink to a stable quiescence radius, as indicated by the evolution of the $FWHM$ and equivalent width of the \ha~line 
 \citep{casares15}. Perhaps,  frequently outbursting  dwarf novae discs always display some level of activity in the form of 
 outflow components \citep{matthews15} that fill-in the line core and prevent unbiased determinations of the trough depth. 
 In any case, more measurements from WZ Sge binaries over a wider range of (accurate) inclinations angles are required to 
 probe whether their accretion discs do follow the $T-i$ correlation of XRTs.

\section{Line profile modeling}
\label{sec:modelling}

In order to investigate  the physical origin of the $T-i$ correlation in XRTs 
we have computed optically thick accretion disc line profiles for a 
range of inclinations\footnote{Note that optically thin lines emit isotropically and, for infinite resolution data, their shape is 
independent of orbital inclination \citep{smak81}.}. 
Following \cite{orosz94}, we model the emission line profile of a flat axisymmetric Keplerian disc using the 
expression 
  
\begin{equation}
F(u) \propto  \int_{r_1}^{r_z} \frac{r^{3/2-\alpha} dr}{\left( 1-u^{2} \, r \right)^{1/2}}  \times \left[ 1+ \left( 2\,  u\,  \sin i \tan i \right) ^{2} r \left( 1-u^{2} \, r \right) \right] ^{1/2} ,
\label{eq:profile}
\end{equation}

\noindent 
where $r$ and $u$ are the dimensionless radius and radial velocity (i.e. normalized to the corresponding values at the outer edge 
of the disc), $r_1$ is the ratio of the inner to the outer disc radius and $r_{z}$=min (1, $u^{-2}$). 
Equation~\ref{eq:profile} assumes that the disc follows a power-law emissivity law of the form $f(r)= r^{-\alpha}$ \citep{smak81},  
with previous works suggesting $\alpha\approx1.5$ \citep{stover81, johnston89, horne91}. 
The inclination-dependent term of the equation accounts for the effect of shear broadening, which 
has been proposed to be the dominant local broadening contribution for  $i\gtrsim60^{\circ}$ \citep{horne-marsh86}.

\begin{figure}
	\includegraphics[angle=-90,width=\columnwidth]{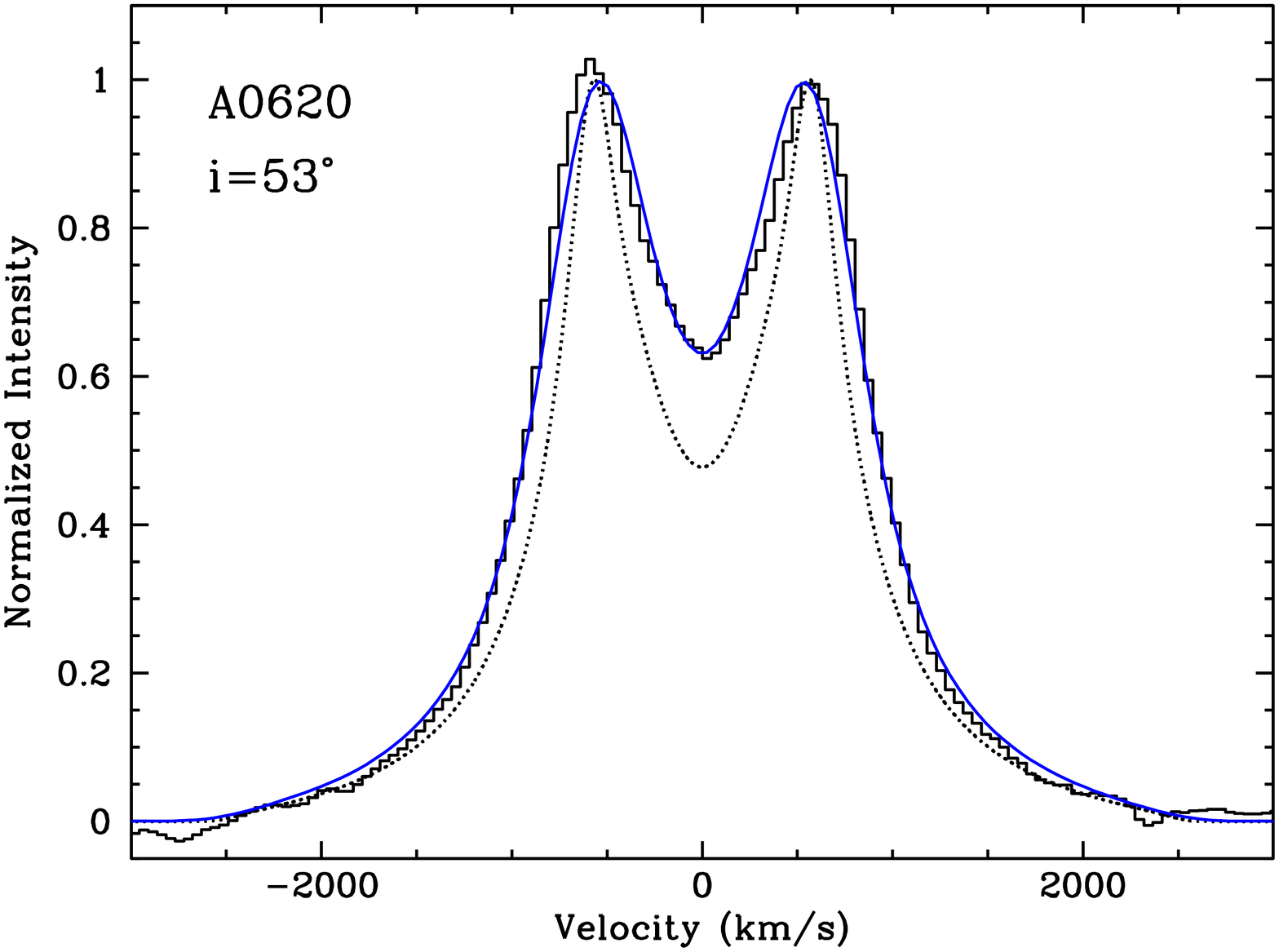}
    \caption{Orbital average GTC \ha~profile of A0620 (black histogram). The black dotted line represents an optically thick model 
    computed from eq.~\ref{eq:profile} for $i=53^{\circ}$, $\alpha=1.3$, $r_{1}=0.05$ and $DP/2=562$ \kms, degraded to an 
    instrumental resolution of 140 \kms. The blue continuous line depicts the same model but further broadened with a 
    Gaussian of $FWHM$=361 \kms.}
    \label{fig:fig5}
\end{figure}

We start by simulating a synthetic profile for the canonical BH XRT A0620, where we adopt    
$i=53^{\circ}$ (Section \ref{sec:calibration}). From our average GTC spectrum we estimate a peak velocity $DP/2=562$ \kms~
and an extreme wing velocity set by the half-width-zero-intensity of the profile i.e. $HWZI=2500$ \kms. Assuming a 
Keplerian disc, with $DP/2$ and $HWZI$ corresponding to the outer and inner disc velocities respectively, we find 
$r_1=(562/2500)^{2}=0.05$.  A synthetic A0620 profile was then computed from eq.~\ref{eq:profile} for 
$i=53^{\circ}$, $r_{1}=0.05$ and $DP/2=562$ \kms. The result was later convolved with a Gaussian of full-width 
at half-maximum $FWHM=140$ \kms~to simulate the instrumental resolution of the 
GTC spectrum. Fig.~\ref{fig:fig5} shows the resolution-degraded profile (dotted line) compared to the GTC data. 
We immediately note that our synthetic model is much narrower than the data and needs to be broadened 
through additional convolution with a Gaussian of $FWHM$=361 \kms~(see Fig.~\ref{fig:fig5}). This quantity was derived  
through a $\chi^2$ minimization process on the residuals obtained after subtracting several broadened versions of the model 
from the observed profile. Stepping through the emissivity law exponent 
we also find that $\alpha=1.3$ provides a good description of the wings in the A0620 profile. 
The excess broadening that we find is not contemplated by eq.~\ref{eq:profile} and reflects some intrinsic (local) broadening 
that we dub $\Delta V_{\rm extra}$. We estimate a typical uncertainty of $\pm$15 \kms~in  $\Delta V_{\rm extra}$ by 
varying $\alpha$ between 1-1.5 and $i$ in the range 50-55$^{\circ}$.   

\begin{figure}
	\includegraphics[angle=-90,width=\columnwidth]{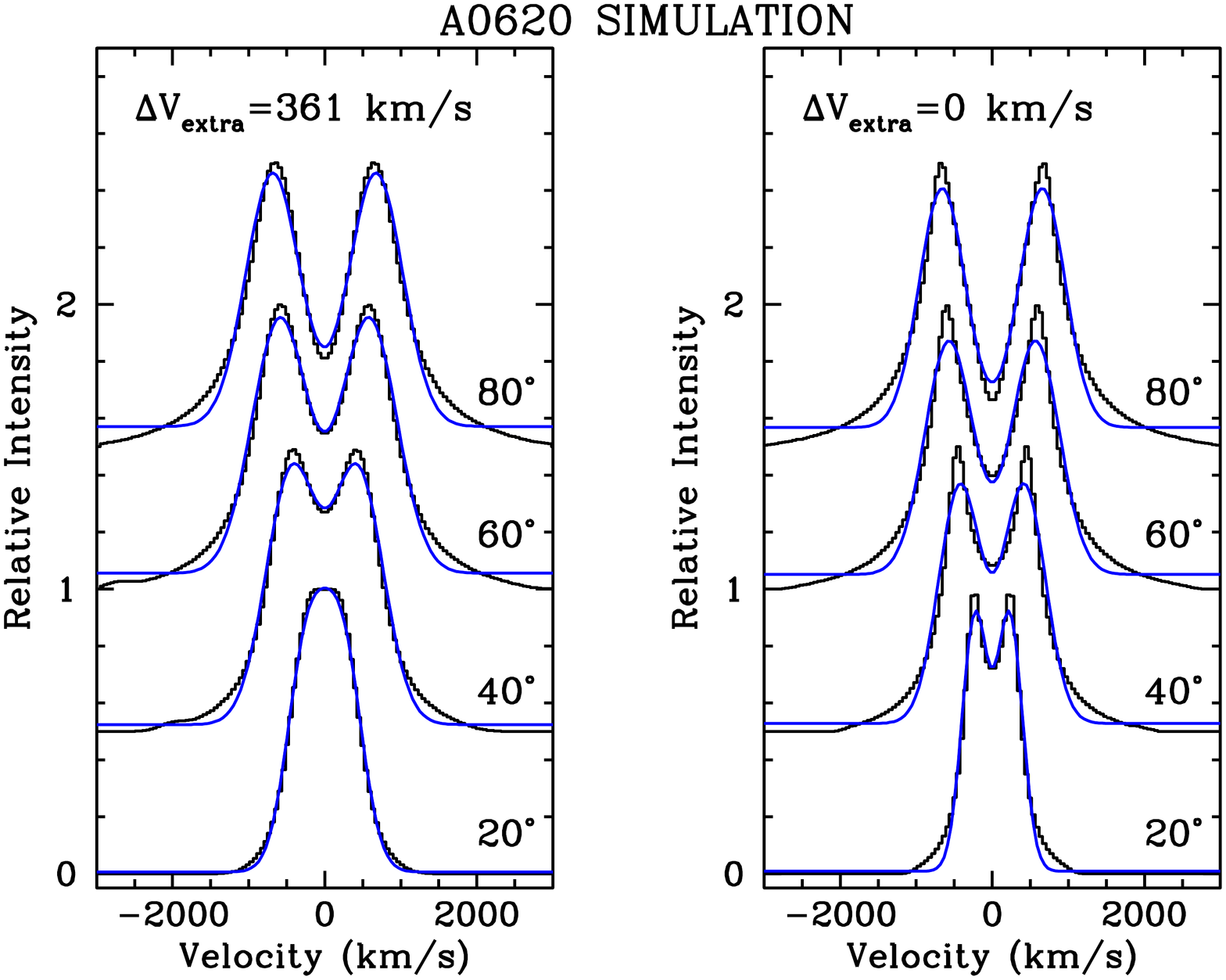}
    \caption{2-Gaussian model fits (blue) to synthetic optically thick disc profiles of A0620 for a range of inclinations. All  
    profiles have been convolved with a Gaussian of 140 \kms~to account for instrumental resolution. Left panel depicts models 
    with added line broadening $\Delta V_{\rm extra}=361$ \kms~while the right panel those 
    with no extra broadening.}
    \label{fig:fig6}
\end{figure}

\begin{figure}
	\includegraphics[angle=0,width=\columnwidth]{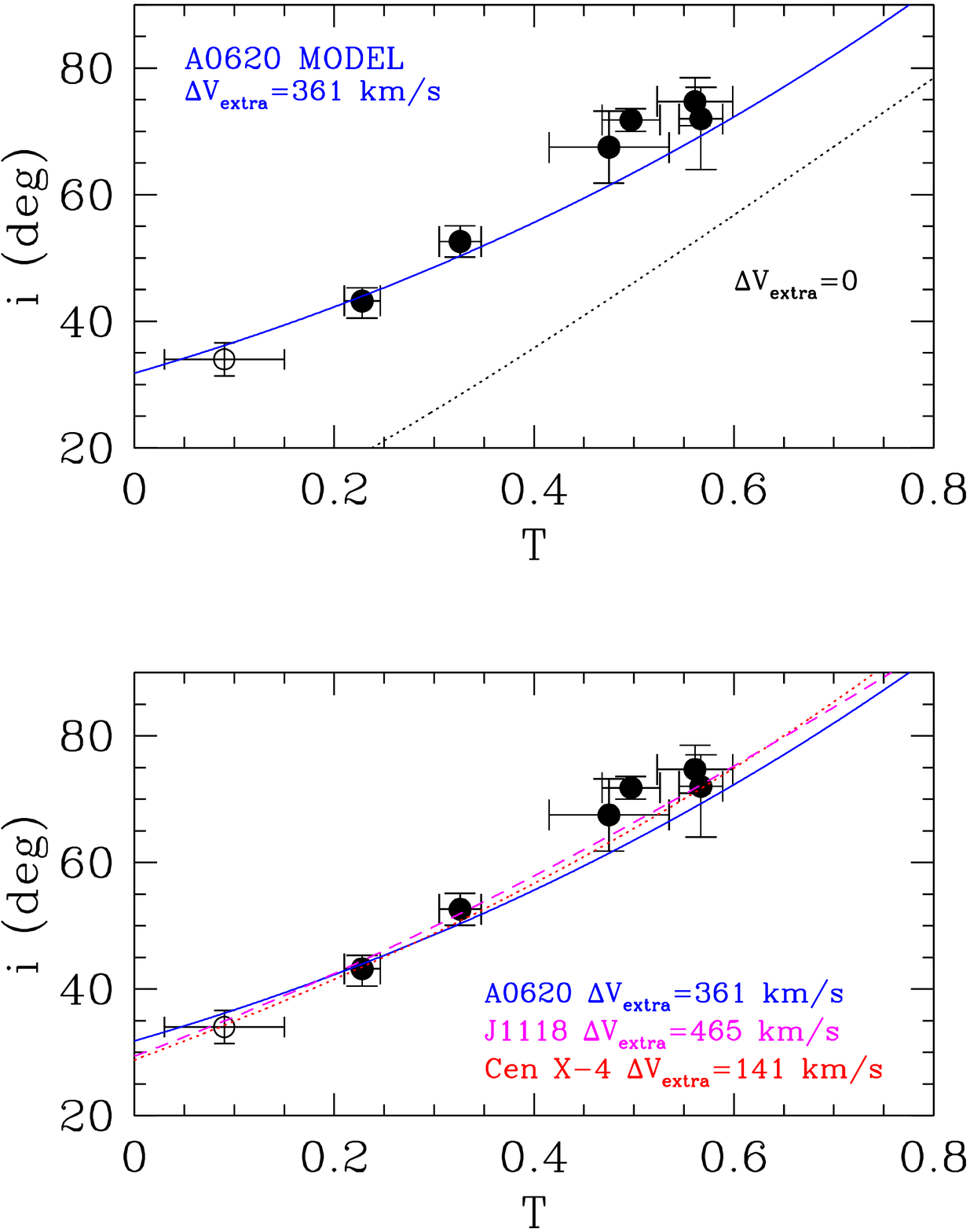}
    \caption{Top: synthetic $T-i$ correlation obtained from an optically thick line model for A0620 with extra 
    broadening $\Delta V_{\rm extra}=361$ \kms (blue line). The black dotted line depicts the same model with 
    $\Delta V_{\rm extra}=0$.  For comparison we overplot the data points of the calibration sample, using the 
    same symbols as in Fig.~\ref{fig:fig2}.  
     Bottom: same for model profiles of Cen X-4 (red dotted line) and J1118 (magenta dashed line), with 
     excess line broadening  $\Delta V_{\rm extra}=141$ and 465 \kms~respectively.}
    \label{fig:fig7}
\end{figure}

We subsequently produced a set of synthetic models with the very same parameters ($\alpha=1.3$, $r_{1}=0.05$, 
$\Delta V_{\rm extra}=361$ \kms) for a range of inclinations between 10$^{\circ}$-90$^{\circ}$ and extracted 
$T$ values by fitting the 2-Gaussian template introduced in Section~\ref{sec:signpost}. 
Note that we here assume that $\Delta V_{\rm extra}$ does not depend on binary inclination. 
For comparison, we also computed synthetic profiles with the same model but adopting $\Delta V_{\rm extra}=0$. 
Fig.~\ref{fig:fig6} displays a selection of simulated profiles and template fits for the two broadening factors, 
while Fig.~\ref{fig:fig7} (top panel) presents the $T-i$ evolution derived from the fits. The latter figure shows that 
the synthetic $T-i$ correlation provides a good description of the empirical data if, and only if, the appropriate 
excess line broadening is taken into account. As a matter of fact, neglecting or underestimating 
$\Delta V_{\rm extra}$ leads to $T$ values that are systematically larger than those predicted by the correlation. 
The figure also evinces an obvious limitation of the technique: the inability to measure low inclinations $i\lesssim30^{\circ}$ 
due to the disappearance of the double peak (see e.g. the $i=20^{\circ}$ profile in the left panel of Fig.~\ref{fig:fig6}). 
This occurs when $W$ approaches $DP$ due to either a high intrinsic broadening $\Delta V_{\rm extra}$, poor spectral 
resolution or both, situation that translates into large values of the {\it scaled-resolution} parameter $\Delta$ (see  Appendix~\ref{ap:resolution}).
 
Since $\Delta V_{\rm extra}$ may vary from system to system we decided to compute
synthetic models for Cen X-4 and J1118 as well. These binaries were selected because 
present us with the narrowest and broadest \ha~profiles within our sample of calibrators. 
As for A0620, optically thick line models were computed for Cen X-4 ($i=34^{\circ}$, $\alpha=1.5$, $r_{1}=0.05$) 
and J1118 ($i=72^{\circ}$, $\alpha=1.3$, $r_{1}=0.12$).  Optimal line broadenings were derived through 
comparison of the resolution-degraded synthetic models with the observed profiles.  These were found to be  
$\Delta V_{\rm extra}=141\pm14$ \kms~and $465\pm40$ \kms~for Cen X-4 and the GTC J1118 spectra, respectively. 
Again, the uncertainties in  $\Delta V_{\rm extra}$ were inferred by varying $\alpha$ in the range 1-1.5 and $i$ between 
the allowed limits (see  Table~\ref{tab:tab1}). Synthetic $T-i$ correlations were subsequently produced by fitting the 
broadened model profiles, computed for a range of  inclinations, with the 2-Gaussian template. The bottom panel 
in Fig.~\ref{fig:fig7} presents the resulting synthetic $T-i$ correlations. 
The plot shows that, despite of the very different $\Delta V_{\rm extra}$ broadenings (141 and 465 
\kms) the two new simulations still provide a good description of the data in the entire range. 

In summary,  we have seen that, notwithstanding obvious limitations at very low inclination angles, the presence of 
some excess broadening,  in addition to line opacity, plays a fundamental role in shaping the line profiles and 
producing the tight $T-i$ correlation that we observe. Every system possesses a characteristic $\Delta V_{\rm extra}$ 
that, if not properly accounted for, would blur the correlation by providing overestimated $T$ values.  

\subsection{Excess broadening}
\label{sec:broadening}

So far we have made the first-order approximation that $\Delta V_{\rm extra}$ is isotropic, i.e. independent from binary inclination. 
Since our calibration database spans a wide range of inclination angles and outer disc velocities it presents us with 
the opportunity to further characterize $\Delta V_{\rm extra}$.  We thus computed synthetic optically thick line models for the 
remaining four XRT calibrators at their corresponding inclinations  and, as before, derived extra line broadenings for each case. 
In order to extend the sample to even larger disc velocities we also included an orbital-averaged GTC spectrum of 
Swift J1357-0933 (hereafter referred to as J1357) from \cite{mata15}, where we adopt $i=87^{\circ}$ (see Section~\ref{sec:2_bhs}). 
In all the cases we find good line fits for $\alpha=1-1.5$ and $r_1$ in the range 0.05-0.13. 
Table~\ref{tab:tab3} lists the optimal excess broadenings obtained for these model profiles ($\Delta V_{\rm extra}$), together with 
their outer disc velocities that we assume close to Keplerian, i.e. $V_{\rm Kep}\approx DP/(2\sin i)$. 
For comparison, we also list the expected values for thermal 
and shear broadening, given by $\Delta V_{\rm th}\sim V_{\rm Kep}~(H/R)$ and 
$\Delta V_{\rm shear}\sim \Delta V_{\rm th}~\sin i~\tan i$ \citep{horne-marsh86}, where  
$R$ and $H$ stand for the outer disc radius and elevation, respectively.  
Here we have adopted a typical thin disc aspect ratio $H/R\sim0.05$.  
In the case of systems with several epochs we have computed $\Delta V_{\rm extra}$ for each epoch individually and 
find very stable values, with typical variations in the range  $\sim10-30$ \kms~and always smaller than the uncertainty 
associated with the choice of model parameters.

\begin{figure}
	\includegraphics[angle=0,width=\columnwidth]{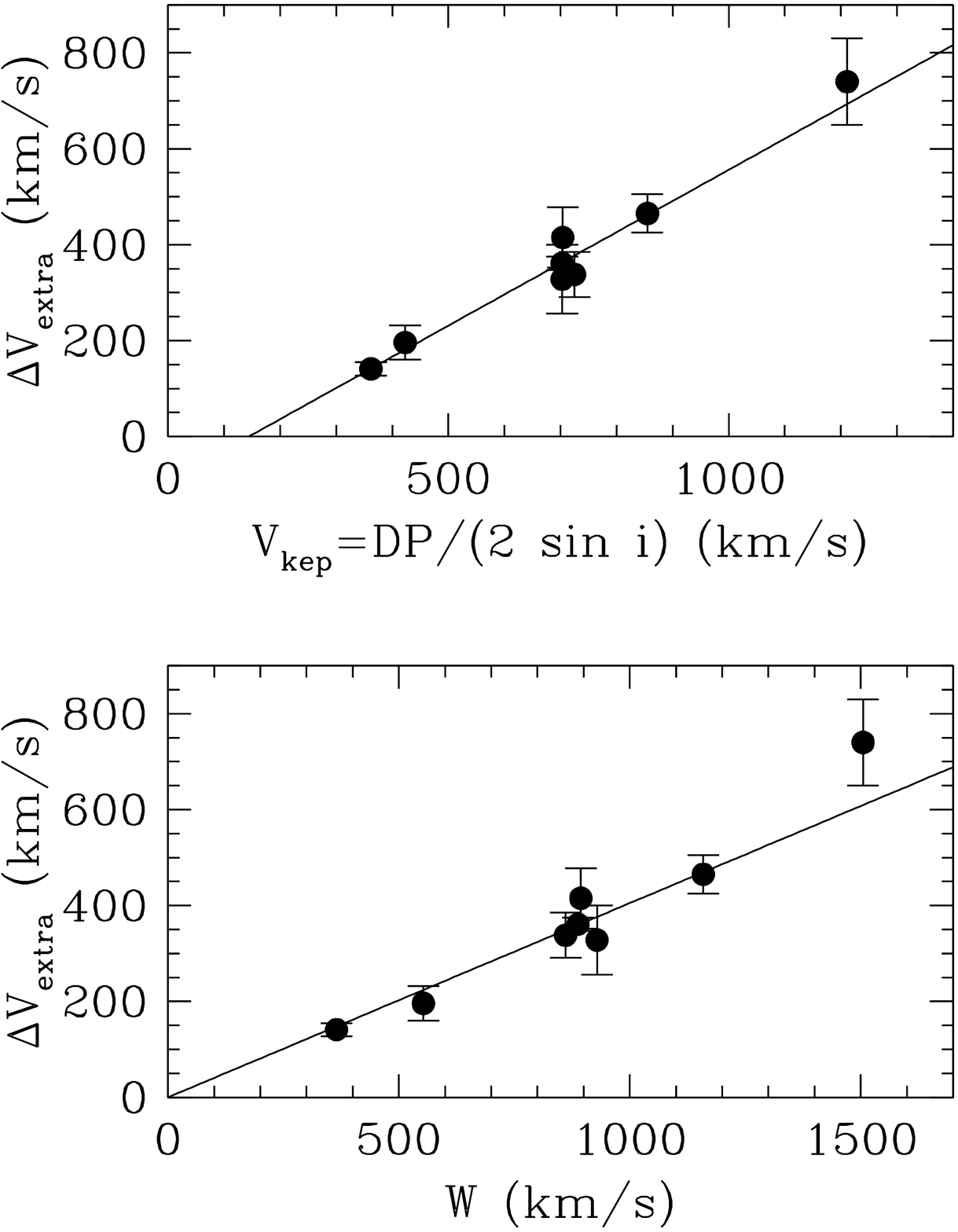}
    \caption{Top panel: Excess line broadening measured in eight XRTs versus deprojected outer disc velocities. The linear fit 
    $\Delta V_{\rm extra}=0.65~V_{\rm Kep}-94$ is also shown. Bottom panel: same versus the intrinsic width of the 2-Gaussian 
    template model. A linear fit  $\Delta V_{\rm extra}=0.41~W$ is overplotted.}
    \label{fig:fig8}
\end{figure}

Table~\ref{tab:tab3} shows that $\Delta V_{\rm extra}$ is much larger than shear broadening, except for J1357. 
The latter is likely a consequence of the $\Delta V_{\rm shear}$ expression breaking down at very extreme 
lines of sight  $\tan i \gtrsim R/H$ (i.e. $i\gtrsim 87^{\circ}$).  
 As a matter of fact, $\Delta V_{\rm extra}$ appears to correlate well with the outer disc velocity (see top panel 
in Fig.~\ref{fig:fig8}) but not binary inclination. In particular, four XRTs with a wide range of inclinations  $i=43-72^{\circ}$ 
(N Mus 91, A0620, GS2000 and J1305) possess very similar $\Delta V_{\rm extra}$ and $
V_{\rm Kep}$ values. Furthermore, J1550, despite its high inclination, has low $\Delta V_{\rm extra}$  and 
$V_{\rm Kep}$ values, comparable to those of the low-inclination binary Cen X-4. A least-squares linear fit to the 
distribution of $\Delta V_{\rm extra}$ and $V_{\rm Kep}$ values yields 

\begin{equation}
\Delta V_{\rm extra}=0.65~V_{\rm Kep}-94. 
\label{eq:blur}
\end{equation}

\noindent
Interestingly, eq.~\ref{eq:blur} offers an alternative route to obtain a rough estimate of the 
binary inclination by simply comparing the double peak separation with $V_{\rm Kep}$. For practical reasons, it is convenient to 
express line broadenings in terms of the $W$ values supplied by the 2-Gaussian template, which are listed in Table ~\ref{tab:tab1}. 
A linear fit (see bottom panel in Fig.~\ref{fig:fig8}) gives\footnote{Note that eq.~\ref{eq:width} encloses the 
assumption that $\Delta V_{\rm extra}$ is the dominant contribution to the intrinsic broadening observed in the line profiles. In reality 
one expects $W=\left( {\Delta V_{\rm extra}}^{2} +  \Delta V_{\rm shear}^{2} + \Delta V_{\rm th}^{2} \right)^{1/2}$ but the good 
correlation seen in the bottom panel of Fig.~\ref{fig:fig6} indicates that neglecting thermal and shear broadening at this stage is 
a fair approximation.}

\begin{equation}
\Delta V_{\rm extra}=0.41~W
\label{eq:width}
\end{equation}

\noindent
and, since  $V_{\rm Kep}\approx DP/(2~\sin i)$, then $ i\approx \arcsin [DP/(289+1.25~W)]$.
Comparing the inclinations derived from this expression with those listed in Table ~\ref{tab:tab1} 
for the seven calibrators we find that the latter can be recovered with a typical uncertainty of $\approx$4 deg.

Furthermore, since $V_{\rm Kep}$ is set by the central object's mass and outer disc radius (which in turn depends 
on the orbital period), $\Delta V_{\rm extra}$ is ultimately determined by fundamental binary parameters. Therefore, 
one can use $W$  to directly constrain the binary period and mass of the compact star. To do so we assume that 
the disc is truncated at the 3:1 resonance radius $R= 3^{-2/3} \left(1+q\right)^{-1/3} a$ \citep{frank02}, where $a$ is 
the binary separation and $q=M_{2}/M_{1}$ the companion star to compact object mass ratio. On the other hand, 
the outer disc velocity is given by  $V_{\rm Kep}=\beta \left( \frac{G M_{1}}{R}\right)^{1/2}$, where $\beta$ is a correction 
factor that accounts for the fact that the outer disc material is sub-Keplerian, due to tidal interaction with the companion 
star. After some algebra, where we bring in Kepler's third Law and eqs.~\ref{eq:blur} and \ref{eq:width}, we find

\begin{equation}
M_{1}^{\ast}=3.45\times10^{-8}~P_{\rm orb}~[\left(0.63~W+145\right)/\beta]^{3}~~~~~M_{\odot}, 
\label{eq:mass}
\end{equation}

\noindent
where $P_{\rm orb}$ is the orbital period in days and $W$ is given in \kms. 
This equation is rather powerful as it allows the measurement of compact object masses 
directly from the width of the line profile, provided that $P_{\rm orb}$ is known (or vice versa). 
And this can be exploited without any prior information on dynamical parameters, the binary mass ratio or  
inclination.  It is worth mentioning that $\Delta V_{\rm extra}$ has been obtained for the \ha~line and, therefore, 
eqs.~\ref{eq:blur}-\ref{eq:mass} should not be applied to other transitions since it has been shown that 
broadening varies between lines (e.g. \citealt{vanspaandonk10}).   

\begin{figure}
	\includegraphics[angle=-90,width=\columnwidth]{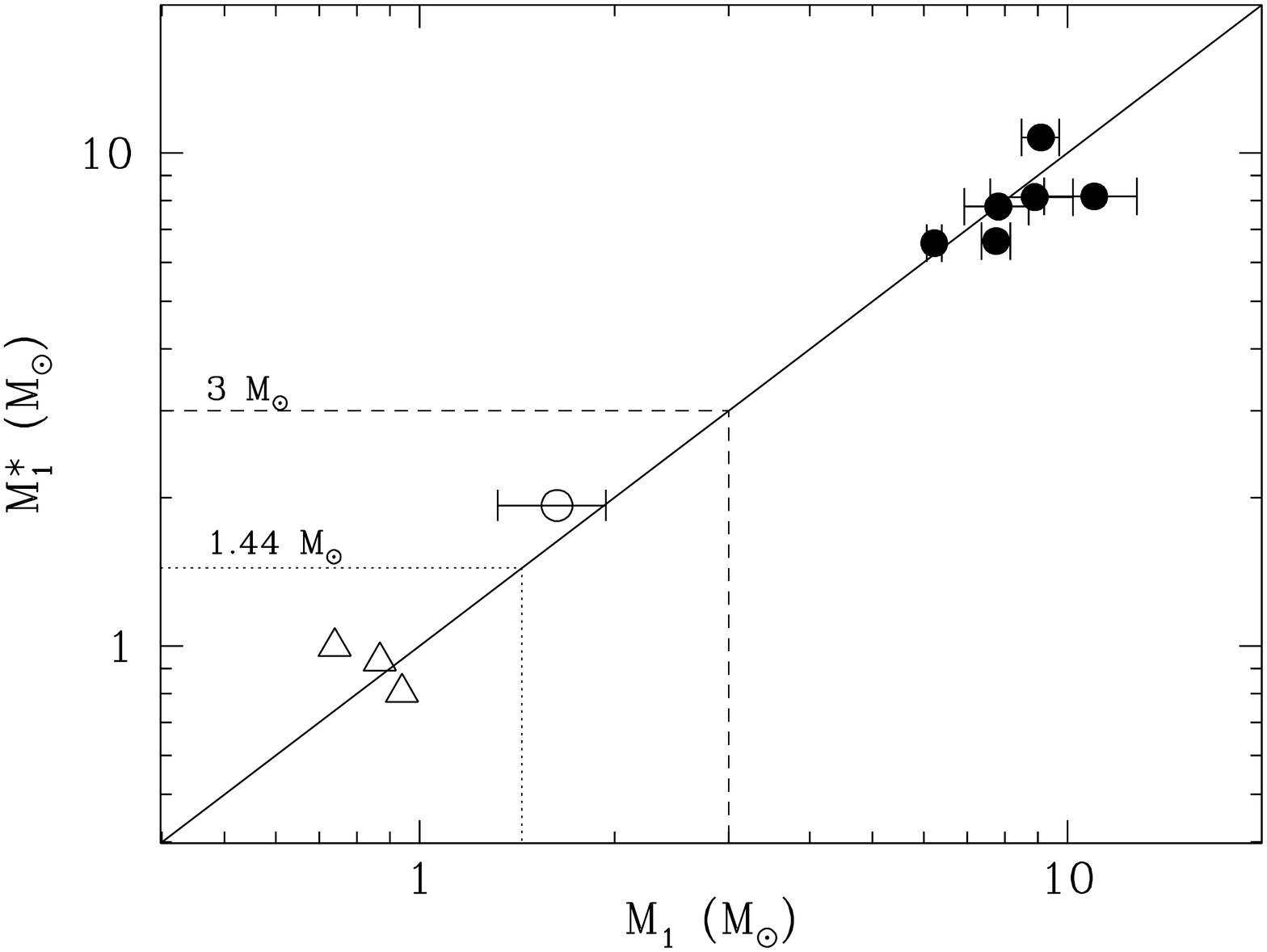}
    \caption{Dynamical masses ($M_{1}$) of the seven calibrators versus masses derived from line profile
     broadenings through eq.~\ref{eq:mass}, with $\beta=0.84$ ($M_{1}^{\ast}$).  We also display the three 
     WZ Sge CVs from Sect.~\ref{sec:cvs}.  Same symbols are used as in Figs.~\ref{fig:fig2} and \ref{fig:fig4}. 
    The maximum NS and Chandrasekhar mass limits are indicated for comparison.} 
    \label{fig:fig9}
\end{figure}

As an example, we have applied eq.~\ref{eq:mass} to our list of calibrators 
and compared the results with dynamical masses reported in literature. As we did for the inclination angle  
(see Section~\ref{sec:calibration}), we here adopt the unweighted mean of the masses reported by independent studies with 
credible inclination determinations. 
For comparison purposes, we have also included the three WZ Sge-type CVs from Section~\ref{sec:cvs}. 
The best match between $M_{1}^{\ast}$ and the dynamical mass $M_1$ 
is obtained for $\beta=0.84$, in good agreement with other studies 
that suggest outer disc velocities are sub-Keplerian by $\approx$20 per cent (e.g. \citealt{wade88,casares16}). 
The results are listed in Table~\ref{tab:tab4} and displayed in Fig.~\ref{fig:fig9}. The distribution of differences 
indicates that eq.~\ref{eq:mass} allows recovering dynamical masses with a $\sim$20 per cent accuracy, 
which is sufficient to separate BHs from other compact stars such as neutron stars (NSs) and white dwarfs. 

To summarize this section, we have shown that optically thick line models can reproduce the observed profiles in XRTs only if 
an extra source of line broadening (surpassing shear broadening) is included. The origin of this supersonic broadening is 
unclear, with Stark broadening \citep{marsh97} or dynamo effects in a magnetically dominated disc \citep{armitage96} as 
possible options. A full discussion of the nature of the excess broadening is beyond the scope of this paper,    
although we note that $\Delta V_{\rm extra}$ increases linearly with the outer disc velocity. 
It is the mere existence of such scaling, together with line opacity, that allows explaining the observed $T-i$ correlation. 
In other words, for a  given binary, the ratio $M_{1}/P_{\rm orb}$ establishes the outer disc velocity $V_{\rm Kep}$ which in turn 
defines both $DP/\sin i$ and $W$, the essential ingredients of the correlation. As a corollary, the scaling between 
excess broadening and outer disc velocity permits deriving fundamental binary 
parameters by simply resolving the width of the double peak emission profile (eq.~\ref{eq:mass}).   
 
\section{Application of the $T-i$ correlation}
\label{sec:2_bhs}

Having presented the $T-i$ correlation and its implications, we can now 
exploit eq.~\ref{eq:correlation} and derive binary inclination angles for BHs with poor or disputed determinations. 
This is mainly the case of binaries where the companion's light curve is heavily veiled/contaminated by strong flickering. 
As an example, we apply the technique to the XRTs J1357 and GRO J0422+32 (=V518 Per; hereafter J0422).

J1357 is characterized by huge flickering variability, with amplitudes up to 1.5 mag that totally conceal 
the companion's ellipsoidal modulation \citep{shahbaz13}. Based on (i) the presence of optical dips, (ii) an extremely 
broad \ha~profile and (iii) a low outburst X-ray luminosity, \cite{corral13} proposed that the source is seen at 
very high (nearly edge-on) inclination $i\gtrsim80^{\circ}$. Further support for an extreme binary geometry was provided by 
observations of deep absorption cores in the \ha~and HeI $\lambda$5876 emission lines \citep{mata15}.  
This interpretation was however disputed by \cite{armas14} and \cite{beri19} through the lack of 
high-inclination spectral and timing X-ray features (see also \citealt{torres15}). 

In the case of J0422, the optical light curves are severely distorted by large flickering variability, with amplitudes 
of up to 0.6 mag. As a result, attempts to model the barely detectable ellipsoidal modulation by different groups have resulted 
in a wide range of inclination angles $i=10-49^{\circ}$ \citep{casares95b, callanan96a, beekman97, webb00, 
gelino03, reynolds07}. 

\begin{figure}
	\includegraphics[angle=0,width=\columnwidth]{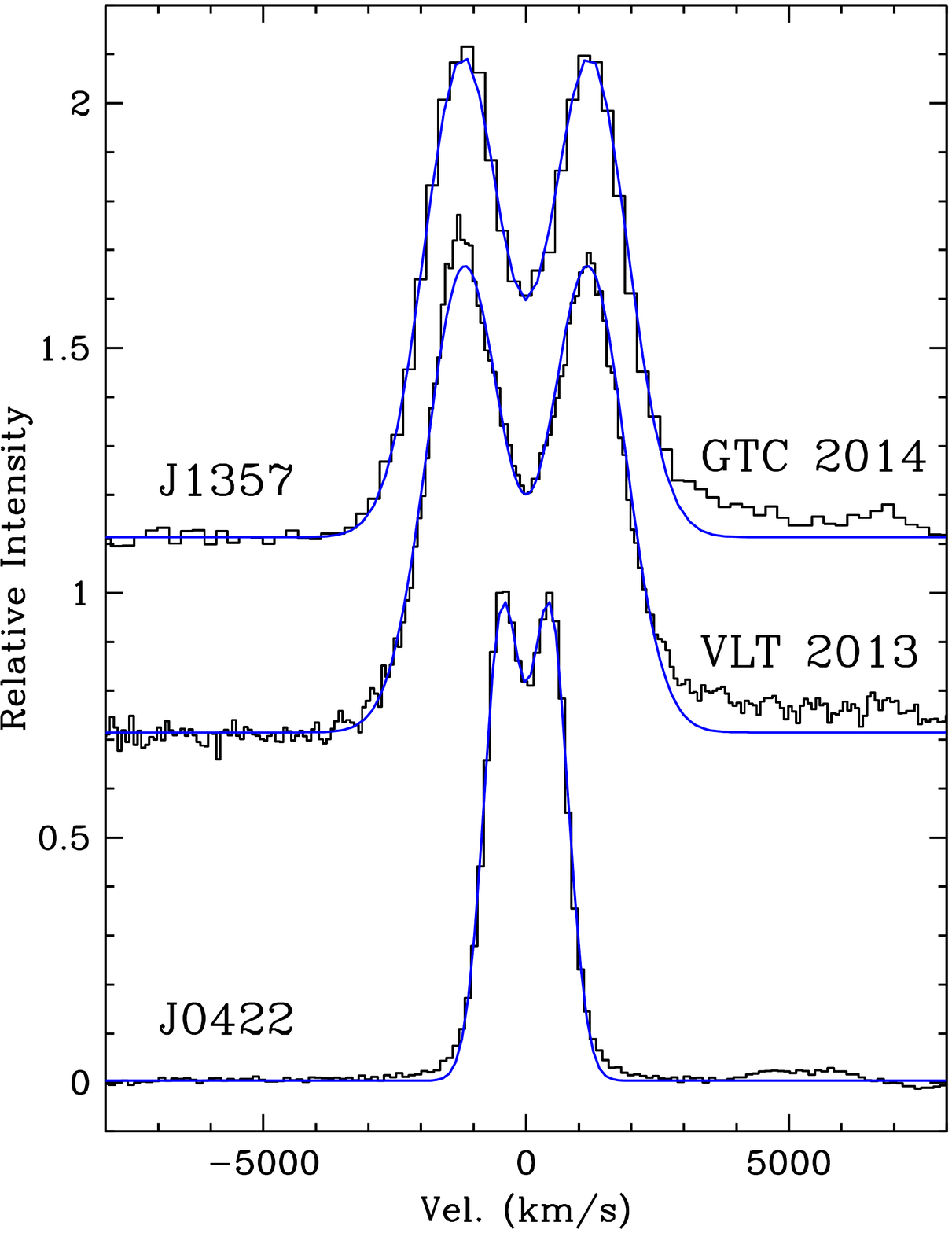}
    \caption{2-Gaussian model fits to orbital averaged spectra of J0422 and two epochs of J1357: VLT 2013 and GTC 2014.  
    The red wing of J1357 (velocities between 3000-8000 \kms) is contaminated by HeI $\lambda$6678 emission 
    and has been masked from the fit.
    } 
    \label{fig:fig10}
\end{figure}

In order to derive \ha~line trough depths we have produced orbital averages for the two binaries. We used 42 GTC spectra of 
J1357 at 800 \kms~resolution, reported in \cite{mata15} and 18 unpublished GTC spectra of J0422, obtained 
in 2016 Jan 9 at 360 \kms~resolution, both covering a complete orbital cycle. 
A thorough analysis of the latter spectra will be presented elsewhere 
(Gonz\'alez Hern\'andez et al. in preparation). Fig.~\ref{fig:fig10} displays the average GTC spectra and 
2-Gaussian model fits for the two systems.  
We find $T=0.683\pm0.006$ and $0.341\pm0.007$ for J1357 and J0422 respectively. 
It should be stressed that, as before, the $T$ values have been obtained by fitting a 2-Gaussian model that has been 
degraded to the instrumental resolution of each database. In addition, because of the 
large {\it scaled-resolution} parameters ($\Delta=0.42$ and 0.69) we apply a small correction 
through eq.~\ref{eq:t_correction}, which translates into +0.017 for J1357 and  +0.046 for J0422. 
Finally, since the $T$ measurements are based on single epoch observations, we decide to adopt 
a typical  uncertainty $\sigma(T)=0.025$ (i.e. a factor 4 larger than the formal error) so to account for the effect of 
possible intrinsic variability (see Sect.~\ref{sec:correlation}). The final fitting results are presented in Table~\ref{tab:tab5}.
The trough depths measured imply $i=87.4^{+2.6}_{-5.6}$ deg and $55.6\pm4.1^{\circ}$ for J1357 and J0422, 
respectively, where the uncertainties have been derived through a Monte Carlo simulation with $10^5$ realizations.  
 
We note that our new J0422 inclination is higher than previous estimates 
(with closest matches provided by \citealt{casares95b} and \citealt{gelino03}) 
although lower than the $63.7\pm5.2^{\circ}$ value inferred by \cite{kreidberg12}, after 
correcting for flickering systematics with a model based on the A0620 disc contamination. 
The inclination of J1357, on the other hand, agrees 
well with \cite{corral13} and \cite{mata21}, thus supporting a near edge-on geometry for this binary. 
We warn, however, that the J1357 inclination should be taken with some caution because it is based on 
extrapolating the $T-i$ correlation beyond the $i=35-75^{\circ}$ domain where it has been tested. 
Even if we were to believe the $T$ values of the WZ Sge stars, J1357 is placed at higher inclinations,    
where the $T-i$ correlation is likely to break down because of extreme lines of sight   
$\tan i \gtrsim R/H$, comparable to the disc aspect ratio. 

As a test, we have also fitted the average of 24 medium resolution (229 \kms) VLT spectra 
of J1357 obtained in 2013 \citep{torres15}. The spectra cover 84 per cent of an orbital cycle. The 2-Gaussian fit yields 
$T=0.503\pm0.025$ which translates into $i=70.6\pm4.7^{\circ}$ (see Fig.~\ref{fig:fig10}). This is significantly 
different than the values obtained from the 2014 GTC spectra.  Most of the discrepancy can be ascribed to the  
larger excess broadening of the VLT spectra, that we estimate to be 1020 \kms. 
Furthermore, we note that the VLT data were collected only 1.6 years after the end of the 2011 outburst, 
compared to 2.8 years for the GTC data. 
Based on the larger excess broadening and close outburst proximity we interpret that the 2013 VLT 
spectra were obtained when the accretion disc did not yet reach complete quiescence and, therefore, give more 
credit to the GTC result. 
The outlier behaviour of the J1357 VLT spectra  is further supported by multi-epoch observations of XTE J1859+226, 
where both $T$ and excess broadening measurements obtained from data collected seven years apart show very 
consistent values, with deviations of only 0.04 and 12 km/s, respectively (Yanes-Rizo et al. 2022, submitted).  
In any case, a more conservative estimate of the inclination in J1357 is provided by the weighted average of $T$ 
values from the two independent epochs, which results in $T=0.612\pm0.125$ and thus $i=81^{+9}_{-12}$ deg.  

\section{Discussion: Updated BH masses and implications}
\label{sec:discussion}

\begin{figure}
	\includegraphics[angle=-90,width=\columnwidth]{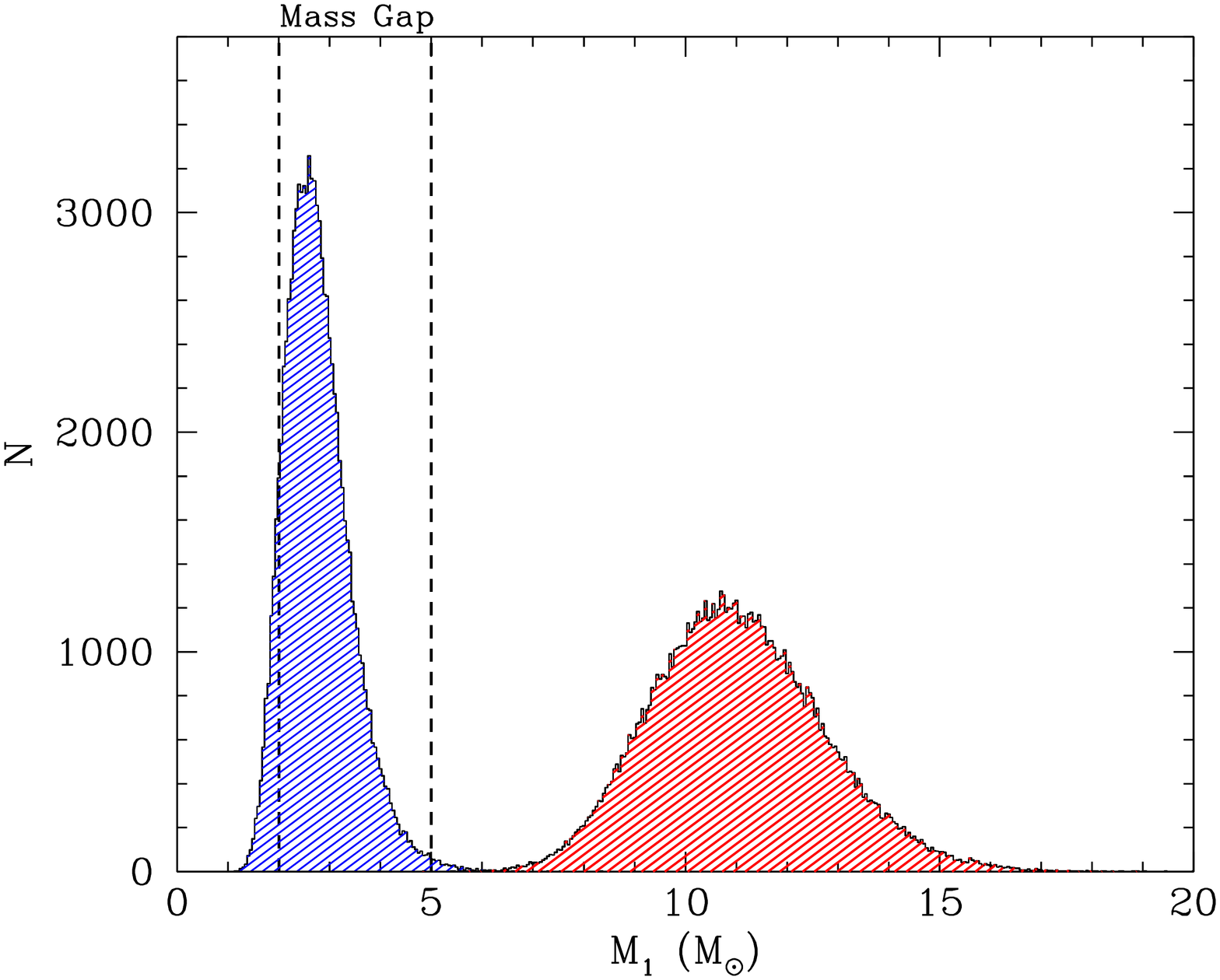}
    \caption{The BH mass probability distributions for J0422 (blue) and J1357 (red) derived from Monte Carlo 
    simulations with 10$^5$ trials. The vertical dashed lines mark the limits of the mass gap between NSs and BHs.} 
    \label{fig:fig11}
\end{figure}

Armed now with the newly derived inclinations we can revise the BH masses in J0422 and J1357 using the observable

\begin{equation}
M_{1}= f(M) \times \frac{(1+q)^2}{\sin^{3} i}
\label{eq:fm}
\end{equation}

\noindent
where $f(M) =  {K_{2}^{3} P_{\rm orb}\over{2 \pi G}}$ is the mass function of the compact star. In order to derive 
$M_{1}$ and its uncertainty we have run Monte Carlo simulations with 10$^5$ realizations. We have adopted 
Gauss-normal distributions for $K_2$, $P_{\rm orb}$ and $q$, as reported in literature, together with our
 recent $i$ determinations (also normally distributed). The adopted values and associated references are 
listed in Table~\ref{tab:tab6}. Fig.~\ref{fig:fig11} presents the probability density functions of the two BH masses, 
with the median values and uncertainties also given in Table~\ref{tab:tab6}.   

Remarkably, we find $M_{1}=2.7^{+0.7}_{-0.5}$ \msun~(68 per cent) for J0422. 
The BH mass lies under 5 \msun~with 99.5 per cent confidence, robustly placing it within the so-called 
 "mass gap" that separates NSs from BHs \citep{bailyn98,ozel10,farr11}. Incidentally, the companion's mass 
 implied by this $M_{1}$ and a well constrained mass ratio (Table~\ref{tab:tab6}) 
 is $M_{2}=q~M_{1}=0.33^{+0.28}_{-0.20}$ \msun, in excellent agreement with the observed M0-5 V spectral type classification 
 \citep{orosz95,casares95b,harlaftis99,webb00}. On the other hand we obtain $M_{1}=10.9^{+1.7}_{-1.6}$ \msun~for J1357 
 ($M_{1}=11.6^{+2.5}_{-1.9}$ \msun~if we adopt a more conservative inclination $i=81^{+9}_{-12}$ deg), 
 which establishes it as one of the most massive BH XRTs in the Galaxy, only comparable to 
 GRS 1915+105 \citep{reid14}\footnote{Note that the BH in the X-ray persistent High Mass X-ray binary Cyg X-1 is even more 
 massive, with  $M_{1}=21.2\pm2.2$ \msun~\citep{miller-jones21}.}. It is interesting to note  the good  agreement between these 
 BH masses and those inferred directly from the width of the 2-Gaussian model fit, coupled with $P_{\rm orb}$ (i.e. $M_{1}^{\ast}$ 
 in Table~\ref{tab:tab4}). 
 
 The evidence of a low-mass BH in J0422 has strong implications for BH formation theories. 
 Several models have been put forward  to explain the apparent dearth of BHs with masses between 2-5 \msun, 
 including fast convection instabilities \citep{belczynski12}, neutrino-driven explosions \citep{ugliano12} or red 
 supergiant failed supernovae \citep{kochanek14}. Other core collapse simulations though, predict a continuous 
 distribution of compact remnants (e.g. \citealt{fryer01}) or the existence of a less populated region under 
 $\approx$6 \msun~but without a real gap \citep{ertl20}. Recent observations 
 of microlense events also favour a continuous distribution of remnant masses \citep{wyrzykowski20}.   
A light BH in J0422 strongly disproves the existence of a mass discontinuity between NSs and BHs, 
lending support to formation scenarios through fallback in weak supernova explosions \citep{ertl20,woosley20} and/or 
accretion-induced collapse of NSs \citep{vietri99,dermer06,gao22}. Further, it suggests that the apparent lack of 2-5 \msun~BHs 
 is not necessarily driven by supernova physics but, instead, could be an artifact of low number statistics, selection biases and 
 possible  systematics associated with the measurement of inclination angles (see \citealt{kreidberg12}). 
 
 Other examples of low-mass BH candidates that may fall into the gap are MWC656 \citep{casares14}, 
 2M05215658+4359220 \citep{thompson19} and V723 Mon \citep{jayasinghe21}, although none presents such a 
 clear-cut case as J0422 (e.g. see BH imposter scenarios in \citealt{vandenheuvel20} and \citealt{el-badry22}).  
Note that, at variance with these other candidates, J0422 was discovered through an accretion-driven outburst 
 and showed X-ray properties that are characteristic of dynamically confirmed BHs, namely 
 (1) a hard power-law spectrum extending beyond 100 keV  \citep{sunyaev93}, (2) a gamma-ray excess at 1-2 MeV 
 \citep{roques94,vandijk95} and (3) the presence of low frequency QPOs \citep{kouveliotou93,vikhlinin95,vanderhooft99}.
   
  Our result on J0422 also impacts the interpretation of gravitational wave (GW) mergers at the low-mass end. 
 For example, GW190425 is believed to result from a binary NS coalescence but its total mass  
 (3.4$^{+0.3}_{-0.1}$ \msun) stands clearly above the mean in the distribution of Galactic binary NSs 
 \citep{abbott20a} and hence the possibility that one of the compact stars was a low-mass BH cannot be 
 dismissed \citep{han20,foley20}. Similarly, the analysis of GW190814 implies the merger of a 22.2-24.3 \msun~BH with a 
 2.50-2.67 \msun~compact star of uncertain nature  \citep{abbott20b}. 
 GW200105 and GW200210, recently disclosed by the GWTC-3 catalogue,  
 also reveal coalescence of BHs with companions of 1.91$^{+0.33}_{-0.24}$ \msun~and 
 2.83$^{+0.47}_{-0.42}$ \msun, that could be either massive NSs or light BHs (see \citealt{abbott21} and references therein).  
In principle, the two scenarios could be distinguished from the imprint of the tidal deformability of the 
NS in the GW signal but unfortunately this will not be possible before third generation detectors 
become operative beyond $\sim$2044 \citep{chen20}.  
Meanwhile, the evidence of a BH in J0422 with a mass of $2.7^{+0.7}_{-0.5}$ \msun~strengthens the possibility that 
low-mass BHs might be present in some GW events.  

\section{Conclusions}
\label{sec:conclusions}

We have shown that the depth of the trough ($T$) in double-peaked \ha~lines from XRT discs 
is linearly correlated with the inclination angle. This is a natural consequence of line opacity and local 
broadening, where a new broadening process (exceeding shear broadening) plays a leading role. 
The here called $T-i$ correlation opens a new avenue to derive binary inclinations 
(and therefore masses) in quiescent XRTs, in particular when ellipsoidal light curves are swamped or 
contaminated by strong flickering. Data with sufficient orbital coverage ($\gtrsim$50 per cent) and spectral resolution 
($\Delta V_{\rm res} \lesssim 0.5 \sqrt{DP^{2}-W^{2}}$) are nevertheless required to prevent $T$ values (and inclinations) 
from being affected by systematic biases.  

We also find that the line excess broadening is proportional to the outer disc velocity.  Such scaling allows inferring 
the ratio $M_{1}/P_{\rm orb}$  directly from the line broadening, without prior knowledge of other fundamental parameters 
(binary mass ratio, companion's velocity or inclination). Therefore, if the orbital period is known, compact object masses can 
be estimated (to $\approx$20 per cent~accuracy) by simply resolving the intrinsic width of the \ha~line (and viceversa). 

We applied the $T-i$ correlation to J0422 and J1357, two BHs with poor inclination constraints due to 
heavy flickering contamination. We find $i=55.6\pm4.1^{\circ}$  and $87.4^{+2.6}_{-5.6}$ deg which lead to  
$M_{1}=2.7^{+0.7}_{-0.5}$ \msun~and  $10.9^{+1.7}_{-1.6}$ \msun~(68 per cent confidence), respectively.  
The BH in J0422 is robustly placed within the 2-5 \msun~mass gap that divides NSs from BHs, 
proving  that low-mass BHs do exist in nature and can be formed through stellar evolution in X-ray binaries. 
This argues for the mass gap being produced by limited statistics and observational biases rather than the physics of core 
collapse supernovae. The result on J0422 also brings new support to the possibility that the light companions in 
the GW190814 and GW200210 mergers (perhaps also GW200105) were low-mass BHs. 
Finally, the $T-i$ correlation presented here opens the possibility to substantially expand the census of 
accurate BH masses and revisit the mass distribution of stellar BHs in the Galaxy. We are currently working 
on this project.

\section{Data availability}
\label{sec:availability}
All data used in this article is publicly available from the relevant observatory archives.  
The data will be shared on reasonable request to the corresponding author. 

\section*{Acknowledgements}

We thank the anonymous referee for carefully reading the manuscript and providing a constructive report that 
has resulted in a significant improvement of the paper.  We are grateful to  D. Steeghs for providing us with the 
Magellan spectra of N Mus 91. Also to  A. Filippenko and J. Orosz for the spectra of GS 2000+25 and 
XTE J1550-564 respectively. Molly software developed  by Tom Marsh is gratefully  acknowledged. 
DMS and MAP acknowledge support from the Consejer\'\i{}a de Econom\'\i{}a, Conocimiento y Empleo del Gobierno de Canarias 
and the European Regional Development Fund (ERDF) under grant with reference ProID2020010104 and ProID2021010132. 
TMD and MAPT acknowledge support via Ram\'on y Cajal Fellowships RYC-2015-18148 and RYC-2015-17854, respectively. 
JIGH acknowledges financial support from the Spanish Ministry of Science and Innovation (MICINN) project PID2020-117493GB-I00.
This work is supported by the Spanish Ministry of Science under grants AYA2017- 83216-P, PID2020-120323GB-I00 and 
 EUR2021-122010.








\begin{landscape}
 \begin{table}
	\centering
	\caption{Calibration binaries}
	\label{tab:tab1}
	\begin{tabular}{lcccccccccc}
		\hline
   Target  & Date & Telescope & $\Delta V_{\rm res}$ & Phase Coverage& $DP$ & $W$  &  $\Delta^{\S}$ & $T$ & $i$ & References \\ 
             & &  & (km/s) & & (km/s) & (km/s) & &   & (deg) & for $i$ \\ 
 		\hline
{\bf Cen X-4}           & 2002-04  & VLT  &    7  & 33\%   & 404.70$\pm$0.03 & 364.92$\pm$0.05 & 0.04 & {\bf 0.090$\pm$0.060}~$^{\dagger}$ & 34.0$\pm$2.6 & 1-3 \\ 
		\hline
{\bf N Mus 91 }   & {\bf Average} &   -      &   -     &     -     &    -     &    -     &      -        &{\bf 0.228$\pm$0.018} & 43.2$^{+2.1}_{-2.7}$ & 4  \\
Epoch 1    & 1993-95 &  NTT   &  90  & 100\% & 989.0$\pm$3.8     & 860.6$\pm$5.4     & 0.18 & 0.199$\pm$0.011 & & \\
Epoch 2    & 2009 &  Magellan  &  46  & 75\% & 1009.4$\pm$2.1     & 855.2$\pm$3.0   &  0.09 & 0.238$\pm$0.006 && \\
Epoch 3    & 2013 &  VLT          &  43  & 70\% & 1015.7$\pm$3.2     & 869.8$\pm$4.5    & 0.08  & 0.223$\pm$0.009 & & \\
		\hline
{\bf A0620 }   & {\bf Average} &   -      &   -     &     -     &    -     &    -     &      -        & {\bf 0.326$\pm$0.021} & 52.6$\pm$2.5  & 5-6  \\
Epoch 1    & 1991-92 & WHT  &    70 & 100\% &1126.6$\pm$0.02 & 893.7$\pm$0.03 & 0.10 & 0.335$\pm$0.001 & & \\
Epoch 2    & 2006 &  Magellan  &  130 & 100\% &997.8$\pm$0.2    & 809.8$\pm$0.2  & 0.19 & 0.302$\pm$0.001 & & \\
Epoch 3   & 2012-13 &  GTC  &  140 & 100\% &1122.9$\pm$0.3    & 887.7$\pm$0.4  & 0.20 & 0.340$\pm$0.001 & & \\
		\hline
{\bf GS2000}   & {\bf Average} &   -      &   -     &     -     &    -     &    -     &      -        & {\bf 0.475$\pm$0.060} & 67.5$\pm$5.7 & 7-8 \\
Epoch 1   & 22/07/1995       & Keck  & 120 & 100\% & 1303.8$\pm$10.0 & 928.5$\pm$12.3   & 0.13 & 0.490$\pm$0.021 &  & \\
Epoch 2   &  25-27/07/1995&  WHT  & 196 & 100\% & 1262.6$\pm$25.4 & 986.5$\pm$34.3   & 0.22 & 0.358$\pm$0.059 & &  \\
		\hline
{\bf J1118}  & {\bf Average} &   -      &   -     &     -     &    -     &    -     &      -        & {\bf 0.497$\pm$0.029} & 71.8$\pm$1.8 & 9-11 \\
Epoch 1  & 2004 & Keck & 50 & 100\% & 1535.9$\pm$1.8   & 1062.8$\pm$2.1   & 0.05 & 0.530$\pm$0.003 &  & \\
Epoch 2 & 07/02/2011 & GTC & 120 & 100\% & 1596.6$\pm$1.3   & 1130.2$\pm$1.6   & 0.11 & 0.498$\pm$0.002 &  & \\
Epoch 3 & 08/02/2011 & GTC & 120 & 100\% & 1557.7$\pm$1.0   & 1126.0$\pm$1.2   & 0.11 & 0.469$\pm$0.002 &  & \\
Epoch 4 & 25/04/2011 & GTC & 120 & 100\% & 1745.1$\pm$1.6   & 1206.8$\pm$1.9   & 0.11 & 0.531$\pm$0.002 &  & \\
Epoch 5 & 12/01/2012 & GTC & 120 & 100\% & 1596.0$\pm$3.2   & 1151.5$\pm$4.0   & 0.11 & 0.472$\pm$0.006 &  & \\
		\hline
{\bf J1305} &  2013 & VLT &140 & 100\% &1339.2$\pm$4.7  & 893.9$\pm$5.9  & 0.14 & {\bf 0.567$\pm$0.022}~$^{\dagger\dagger}$ & 72.0$^{+5.0}_{-8.0}$ & 12 \\
		\hline
{\bf J1550}  & 2008  & Magellan & 55 & 30\% & 817.3$\pm$3.7  & 552.5$\pm$4.3 &  0.09 & {\bf 0.561$\pm$0.038}~$^{\dagger\dagger\dagger}$ & 74.7$\pm$3.8 & 13 
\\
		\hline
	\end{tabular}
\end{table}
{\bf References:} (1)  \cite{khargharia10}, (2) \cite{shahbaz14}, (3) \cite{hammerstein18}, (4) \cite{wu16}, 
(5) \cite{cantrell10}, (6) \cite{vangrunsven17}, (7) \cite{callanan96b}, (8) \cite{ioannou04}, (9) \cite{gelino06}, 
(10) \cite{khargharia13}, (11) \cite{cherepashchuk19}, (12) \cite{mata21} and (13) \cite{orosz11} \\ 
$^{\S}$~{\it Scaled-resolution} parameter, defined as $\Delta=\Delta V_{\rm res}/\sqrt{DP^{2}-W^{2}}$. \\
$^{\dagger}$~$T$ value and error corrected for orbital phasing.\\
$^{\dagger\dagger}$~$T$ value corrected by -0.012 and formal error increased by +0.012 to account for masking the 
contamination of an interloper star in the line core.\\
$^{\dagger\dagger\dagger}$~An overall error of 0.038 has been adopted to account for the possible effect of orbital variability.
\end{landscape}


\begin{landscape}
 \begin{table}
	\centering
	\caption{Cataclysmic variables}
	\label{tab:tab2}
	\begin{tabular}{lccccccccc}
		\hline
   Target  & CV type & $\Delta V_{\rm res}$ & Phase Coverage& $DP$ & $W$  &  $\Delta$ & $T$ & $i$ & References \\ 
         &   & (km/s) & & (km/s) & (km/s) & &   & (deg) & for spectra \& $i$ \\ 
 		\hline
U Gem    &  Dwarf novae  & 16   & 100\%   &   908.7$\pm$1.0   & 726.0$\pm$1.3   &  0.03  &  0.325 $\pm$0.003 & 69.7$\pm$ 0.7 & 1,2 \\ 
HT Cas   &  ~~~,,   &120  &   60\%   & 1089.4$\pm$2.9   &  922.3$\pm$4.0  &  0.21  & 0.230$\pm$0.007 & 81.0$\pm$1.0  & 3,4  \\
OY Car   &    ~~~,,   & 46   & 100\%   & 1296.6$\pm$0.3   & 942.2$\pm$0.4   &  0.05  & 0.462$\pm$0.001 & 83.3$\pm$0.2  & 5,6 \\
IP Peg    &    ~~~,,   &150  & 100\%   &  1066.7$\pm$4.6  &  899.9$\pm$6.5  & 0.26  &  0.245$\pm$0.012 & 83.8$\pm$0.2  & 7,8  \\
J1300     &     ~~~,,   &45   & 100\%   & 1164.6$\pm$0.7   & 900.7$\pm$0.9.  & 0.06  & 0.372$\pm$0.002 & 85.7$\pm$1.5 & 9 \\
		\hline
WZ Sge  &  WZ Sge & 25   & 100\% & 1223.9$\pm$0.6   & 835.2$\pm$0.7 & 0.03 & 0.549$\pm$0.001 &  75.9$\pm$0.3 & 10,11 \\
J1035      &   ~~~,,   & 69   & 100\% & 1254.8$\pm$1.5   & 809.6$\pm$1.7 & 0.07 & 0.622$\pm$0.003 &  83.1$\pm$0.2 & 6,12 \\
J1433      &  ~~~,,   &  32   & 100\% & 1368.5$\pm$2.4   & 826.9$\pm$2.7 &  0.03 & 0.700$\pm$0.004 & 84.2$\pm$0.2 & 6,13 \\
\hline
	\end{tabular}
\end{table}
{\bf References:} (1) \cite{echevarria07}, (2)  \cite{zhang87}, (3) \cite{casares15}, (4)  \cite{horne91b}, (5) \cite{copperwheat12}
(6) \cite{littlefair08}, (7) \cite{harlaftis94}, (8) \cite{copperwheat10}, (9) \cite{savoury12}, (10) \cite{skidmore00}, (11) \cite{skidmore02}, 
(12) \cite{southworth06}, (13) \cite{tulloch09}  \\
\end{landscape}


\begin{landscape}
 \begin{table}
	\centering
	\caption{Sources of local line broadening in 8 XRTs}
	\label{tab:tab3}
	\begin{tabular}{lccccc}
		\hline
   Target  & $i$ & $V_{\rm Kep}$ & $\Delta V_{\rm extra}$  & $\Delta V_{\rm th}$ & $\Delta V_{\rm shear}$ \\ 
             & (deg) & (km/s) & (km/s) & (km/s) &  (km/s)  \\ 
 		\hline
Cen X-4               & 34  & 362  & 141$\pm$14 & 18 & 7 \\ 
N Mus 91             &  43 &  725 & 338$\pm$47 & 36 & 23 \\
A0620             &  53  &  703 & 361$\pm$15 & 35 & 37 \\
GS2000        &   68 &  703  & 328$\pm$72 & 35 & 80 \\
J1118   &   72 &  855  & 465$\pm$40 & 43 & 125 \\
J1305  &   72 &  704  & 415$\pm$63 & 35 & 103 \\
J1550    &   75 &  423  & 196$\pm$36 & 21 & 76 \\
J1357  & 87 & 1211 & 740$\pm$90 &  61 & 1089  \\
\hline
	\end{tabular}
\end{table}
\end{landscape}

\begin{landscape}
 \begin{table}
	\centering
	\caption{Compact objects masses from line broadenings ($M_{1}^{\ast}$) compared to dynamical masses ($M_{1}$)}
	\label{tab:tab4}
	\begin{tabular}{lccccc}
		\hline
   Target  & $P_{\rm orb}$ & $W$ & $M_{1}^{\ast}$  & $M_{1}$ & References \\ 
               & (d) & (km/s) &      (\msun)$^{\dagger}$ &   (\msun) & for $M_{1}$ \\ 
 		\hline
Cen X-4               &  0.6290522(4)       &   364.92$\pm$0.05  & 1.93$\pm$0.001 & 1.64$\pm$0.31 & 1-3 \\ 
N Mus 91             &  0.43260249(9)     &   860.6$\pm$5.4      &  8.16$\pm$0.12 & 11.0$^{+2.1}_{-1.4}$ & 4 \\
A0620             &    0.32301405(1)   &  887.7$\pm$0.4      &  6.56$\pm$0.01 &  6.23$\pm$0.17  & 5-6 \\
GS2000        &     0.3440915(9)    &   928.5$\pm$12.3    & 7.78$\pm$0.24 & 7.82$\pm$0.89 & 7-8 \\
J1118   &     0.16993404(5)  &   1159.2$\pm$0.9    & 6.63$\pm$0.01 &  7.76$\pm$0.40 & 9-11 \\
J1305  &     0.394(4)           &   893.9$\pm$5.9       & 8.14$\pm$0.12 & 8.9$^{+1.6}_{-1.0}$ & 12 \\
J1550    &    1.5420333(24)  & 552.5$\pm$4.3        & 10.75$\pm$0.17 & 9.10$\pm$0.61 & 13 \\
		\hline
WZ Sge  &  0.0566878460(3) &  835.2$\pm$0.7  & 0.997$\pm$0.002 &  0.740$\pm$0.071 &  14 \\
J1035 &   0.0570067(2)      &   760.0$\pm$1.2  & 0.805$\pm$0.003 & 0.94$\pm$0.01 & 15  \\
J1433 &   0.054240679(2)  &     826.9$\pm$2.7 & 0.931$\pm$0.007 & 0.868$\pm$0.007 & 15 \\
\hline
	\end{tabular}
\end{table}
{\bf References:} (1)  \cite{khargharia10}, (2) \cite{shahbaz14}, (3) \cite{hammerstein18}, (4) \cite{wu16}, 
(5) \cite{cantrell10}, (6) \cite{vangrunsven17}, (7) \cite{callanan96b}, (8) \cite{ioannou04}, (9) \cite{gelino06}, 
(10) \cite{khargharia13}, (11) \cite{cherepashchuk19}, (12) \cite{mata21}, (13) \cite{orosz11},  
(14) \cite{skidmore02}, (15) \cite{littlefair08}\\
$^{\dagger}$~$M_{1}^{\ast}$ is obtained from eq.~\ref{eq:mass}, with $\beta=0.84$. Quoted errors are pure 
formal uncertainties.\\
\end{landscape}
\begin{landscape}
 \begin{table}
	\centering
	\caption{Profile fit parameters and inclinations in 2 BH XRTs}
	\label{tab:tab5}
	\begin{tabular}{lccccccc}
		\hline
   Target  & $\Delta V_{\rm res}$ & Phase Coverage& $DP$ & $W$ & $\Delta$ & $T^{\dagger}$ & $i$ \\ 
               & (km/s) & & (km/s) & (km/s) &   & & (deg) \\ 
 		\hline
J1357           &  800 & 100\% & 2418.5$\pm$4.6 & 1504.6$\pm$6.4 & 0.42 & 0.683$\pm$0.025 &  87.4$^{+2.6}_{-5.6}$  \\ 
J0422            &  360 & 100\% &  908.0$\pm$2.1  &   740.5$\pm$3.3  & 0.69 & 0.341$\pm$0.025 & 55.6$\pm$4.1  \\
\hline
	\end{tabular}
\end{table}
$^{\dagger}$~Because of the large $\Delta$ values the quoted $T$ have been corrected for instrument resolution 
 degradation using eq.~\ref{eq:t_correction}. 
\end{landscape}


\begin{landscape}
 \begin{table}
	\centering
	\caption{Fundamental parameters and dynamical masses for 2 BH XRTs}
	\label{tab:tab6}
	\begin{tabular}{lccccccc}
		\hline
   Target  & $P_{\rm orb}$ & $K_2$ & $q$ & $i$ & $M_{1}$  & $M_{1}^{\ast}$ & References \\ 
               & (d) & (km/s) &   & (deg) &   (\msun) &   (\msun)$^{\dagger}$ &  \\ 
 		\hline
J1357  & 0.106969(23) & 967$\pm$49  & 0.039$\pm$0.004 & 87.4$^{+2.6}_{-5.6}$ & 10.9$^{+1.7}_{-1.6}$  & 8.1$\pm$1.6 & 1-3 \\ 
J0422  & 0.2121600(2) & 378$\pm$16  & 0.12$\pm$0.08  & 55.6$\pm$4.1 & 2.7$^{+0.7}_{-0.5}$ & 2.8$\pm$0.6 & 1, 4-5  \\
\hline
	\end{tabular}
\end{table}
{\bf References:} (1) this paper, (2) \cite{mata15}, (3) \cite{casares16}, (4) \cite{webb00} and (5) \cite{harlaftis99}. \\
$^{\dagger}$~$M_{1}^{\ast}$ is obtained from eq.~\ref{eq:mass},  with $\beta=0.84$. Quoted errors are 20 per cent. \\
\end{landscape}


\appendix

\section{Impact of instrumental resolution in line trough measurements}
\label{ap:resolution}

We here investigate the impact of instrumental resolution in $T$  measurements. Line profiles observed at low resolution will 
appear blurred and, therefore, the depth of the central depression will be underestimated. Since $T$ is determined by the ratio 
$(DP/W)$ the accuracy in the final $T$ value is driven by how instrumental resolution ($\Delta V_{\rm res}$) compares with 
both, the double peak separation and intrinsic line broadening. After several tests 
we find that the quantity $\Delta=\Delta V_{\rm res}/\sqrt{DP^{2}-W^{2}}$  provides a good figure of merit to 
assess how accurately $T$ can be recovered for a given instrumental resolution. We name this quantity the 
{\it scaled-resolution} parameter. 

To illustrate the effect of instrument resolution in the fitted quantities we have simulated three double-peaked 
profiles mimicking the cases of Cen X-4 ($DP_0$=405, $W_0$=365 \kms), A0620 ($DP_0$=1125, $W_0$=890 \kms) 
and J1118 ($DP_0$=1625, $W_0$=1160 \kms).  We here use the subindex "0" to indicate the initial quantities, before 
degradation by instrument resolution. The three profiles encompass a sufficiently wide range of parameters so to 
represent most observed cases. We subsequently convolved these profiles with Gaussian functions of increasing 
width in the range 50-1150 \kms~to simulate instrument resolution $\Delta V_{\rm res}$  and 
performed 2-Gaussian model fits to the results. The profiles were sampled with a pixel size of $1/4\times \Delta V_{\rm res}$ 
and noise was added to represent a typical observation.

\begin{figure}
	\includegraphics[angle=0,width=\columnwidth]{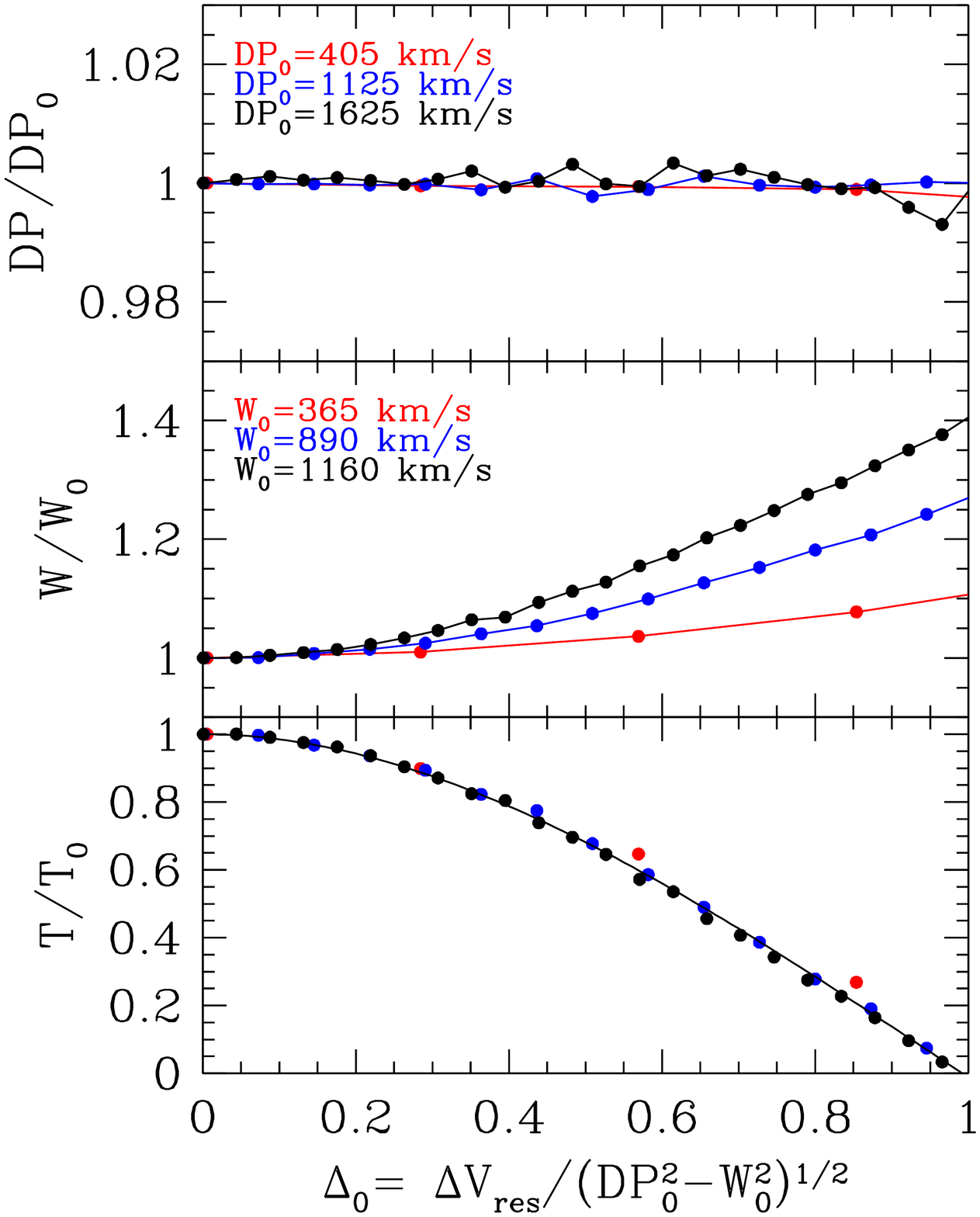}
    \caption{
    Impact of instrumental resolution on  double-peaked profile parameters. The three panels show the 
    evolution of  $DP$, $W$ and $T$ versus the {\it scaled-resolution} parameter $\Delta_{0}$.  
    $DP$, $W$ and $T$ are obtained by performing plain 2-Gaussian model fits to progressively degraded versions 
    of synthetic double-peaked profiles that simulate the cases of Cen X-4 (red), A0620 (blue) and J1118 (black). 
    The line profile parameters are plotted normalized to the initial (non-degraded) values $DP_{0}$, $W_{0}$ and $T_{0}$. 
     The curved line in the bottom panel represents a cubic fit to the ensemble of points. 
}
    \label{fig:a1}
\end{figure}

Fig.~\ref{fig:a1} depicts the evolution of the fitted parameters $DP$, $W$ and $T$  as a function of the 
{\it scaled-resolution} $\Delta_{0}$ for the three cases investigated.  
The parameters are shown normalized to their original non-degraded values. 
Interestingly, we observe that $DP$ is little affected by instrument resolution 
since it is measured to within $\sim$0.5 per cent of its initial value, even at poor resolutions (i.e. large $\Delta_{0}$ values). 
$W$, on the other hand, rises with $\Delta_{0}$ and, consequently, $T$ declines. 
In the limit when  $\Delta V_{\rm res}=\sqrt{DP_{0}^{2}-W_{0}^{2}}$ (i.e. $\Delta_{0}=1$) the double peak trough vanishes 
since $T=0$. We find that the evolution of $T$ with instrumental resolution is well described by the cubic expression 
$T/T_{0}= 1+(0.53 \Delta_{0}-1.54)  \Delta_{0}^{2}$.  

\begin{figure}
	\includegraphics[angle=0,width=\columnwidth]{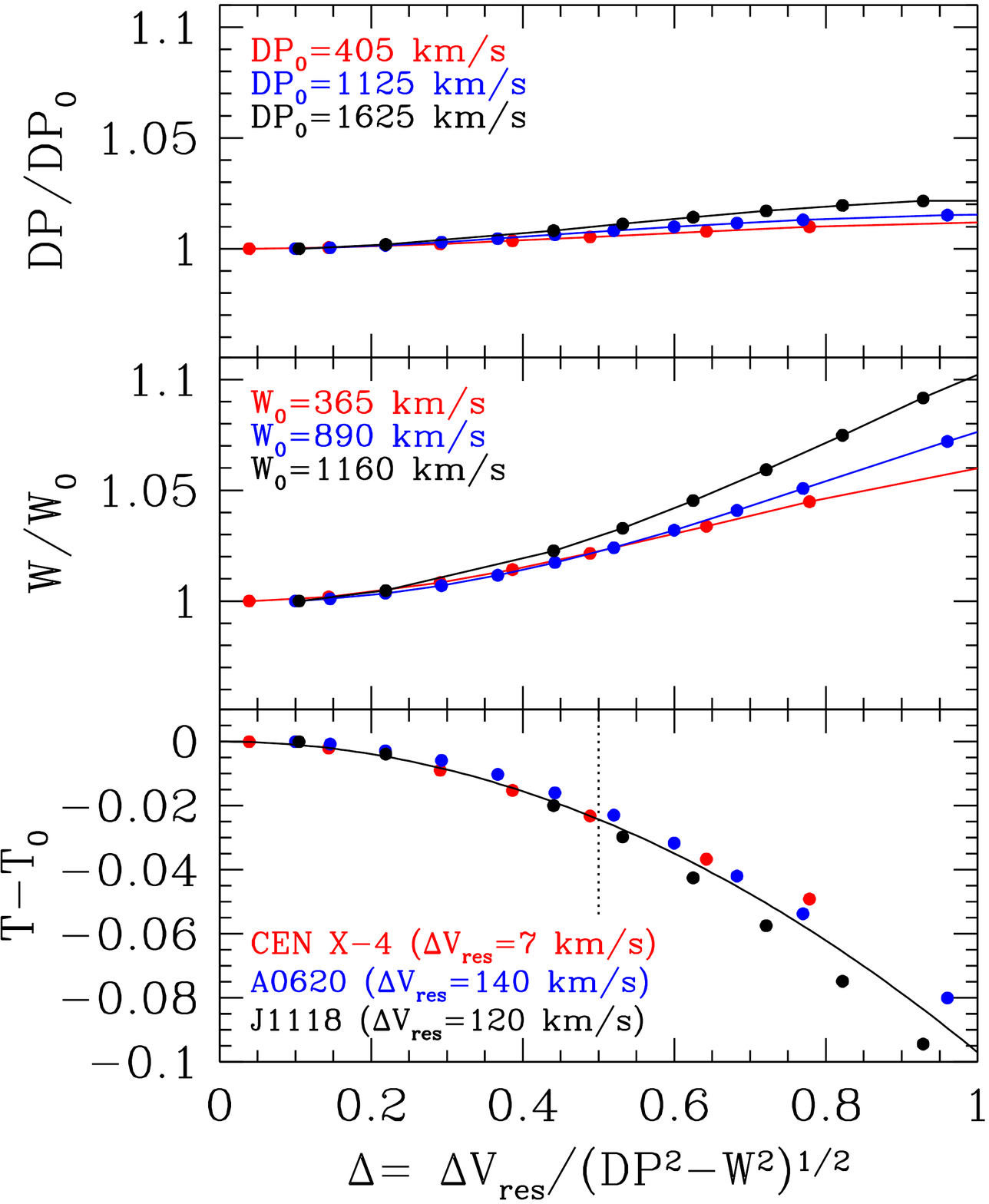}
    \caption{
    Deviation between the original line parameters ($DP_0$, $W_0$ and $T_0$) and those inferred by fitting 
    resolution-degraded 2 Gaussian models. This simulation employs real Cen X-4,  J1118 and A0620 
    spectra, progressively degraded with increasing resolutions. The curved solid line in the bottom panel represents a quadratic 
    fit to the ensemble of points (eq.~\ref{eq:t_correction}). The vertical dotted line marks a reference limit above which 
    instrumental resolution starts to have a significant impact on $T$ values extracted from degraded models.   
    In other words, for $\Delta > 0.5$ the measured $T$ values are underestimated by more than the typical variability 
    $\sigma (T) \approx 0.025$ observed in multi-epoch observations.   
    }
    \label{fig:a2}
\end{figure}

Of course, when dealing with real data the intrinsic line parameters $DP_{0}$ and $W_0$ (and thus $\Delta_{0}$) are not 
known beforehand. In principle, these can be inferred by fitting 2-Gaussian models that are previously 
degraded to the resolution of the data (i.e. by convolution with a Gaussian of $FWHM=\Delta V_{\rm res}$) and this 
is the approach that we have followed throughout the paper. In order to test  how close the inferred $DP$ and $W$ values 
are to the intrinsic quantities we have performed a simulation using real data.   
We have taken the average spectra of Cen X-4 plus the GTC spectra of A0620 and J1118 (see Section 3), and degraded 
them further through convolution with Gaussian functions of increasing widths. The effective resolution of the new degraded 
profiles will thus be the quadratic sum of the instrument resolution (i.e. 7, 140 and 120 \kms~for Cen X-4, A0620 and J1118, 
respectively) and the Gaussian convolution widths. We then performed 2-Gaussian model fits, degraded by their 
corresponding effective resolutions, to the new profiles. The results are plotted as a function of the scaled-resolution parameter 
in Fig.~\ref{fig:a2}.  Note that here we plot $\Delta=\Delta V_{\rm res}/\sqrt{DP^{2}-W^{2}}$ in the x-axis, where 
$DP$ and $W$ represent the  inferred $DP_0$ and $W_0$ values, as derived from the fits. Since the initial instrumental 
resolution in the three binaries is sufficiently small we take the $DP$, $W$ and $T$ values measured from
the original (non-degraded) profiles as the true unbiased quantities i.e. $DP_{0}$, $W_{0}$ and $T_{0}$.

The bottom panel of Fig.~\ref{fig:a2} shows that the $T$ values provided by the resolution-degraded model fits are still 
underestimated, but to a much lesser extent than before. In order to quantify the systematic shift in $T$ measurements
we have fitted a quadratic function to the bottom panel of  Fig.~\ref{fig:a2} and 
find

\begin{equation}
T-T_{0}= -0.097 \Delta^{2}. 
\label{eq:t_correction}
\end{equation}

\noindent
We observe that for $\Delta\leq0.2$ (i.e. the case of our seven calibrators) the original $T$ values are accurately recovered 
to better than 0.005. As the scaled-resolution parameter increases, though, the measured $T$ values become more 
and more biased low. Still, for $\Delta<0.5$ the $T$ values are underestimated by less than the typical 
variability observed in multi-epoch observations (i.e. $\sigma (T) \simeq 0.025$) and, therefore, we conclude 
they are not significantly biased by instrumental resolution. 
Anyhow eq.~\ref{eq:t_correction} provides a way to correct for the systematic shift in $T$ measurements 
from resolution-degraded models that we believe is  reliable up to $\Delta \approx 0.8$. 
For example, in the cases of J1357 and J0422, with $\Delta=0.42$ and 0.69, we estimate a correction of 
+0.017 and +0.046, respectively (Sect.~\ref{sec:2_bhs}). 
 
\section{Orbital dependence of line trough}
\label{ap:orbital}

The measurement of the line trough $T$  is best performed over orbital averages because (1) individual spectra rarely 
possess enough signal-to-noise for the fitting technique to be applicable and (2) orbital means average out 
possible asymmetries in individual spectra from, for example, hot-spots or disc eccentricities which could potentially 
bias the determination of $T$.  
We here explore the impact of limited phase coverage in $T$ measurements by fitting our 2-Gaussian model to a sample 
of systems with high quality phase-resolved individual spectra. These are the GTC spectra of J0422 
(Section \ref{sec:2_bhs}), A0620, J1118 and J1357, plus a VLT database on J1357 
from \cite{torres15}. The results are presented in Fig.~\ref{fig:a2}.  

\begin{figure}
	\includegraphics[angle=-90,width=\columnwidth]{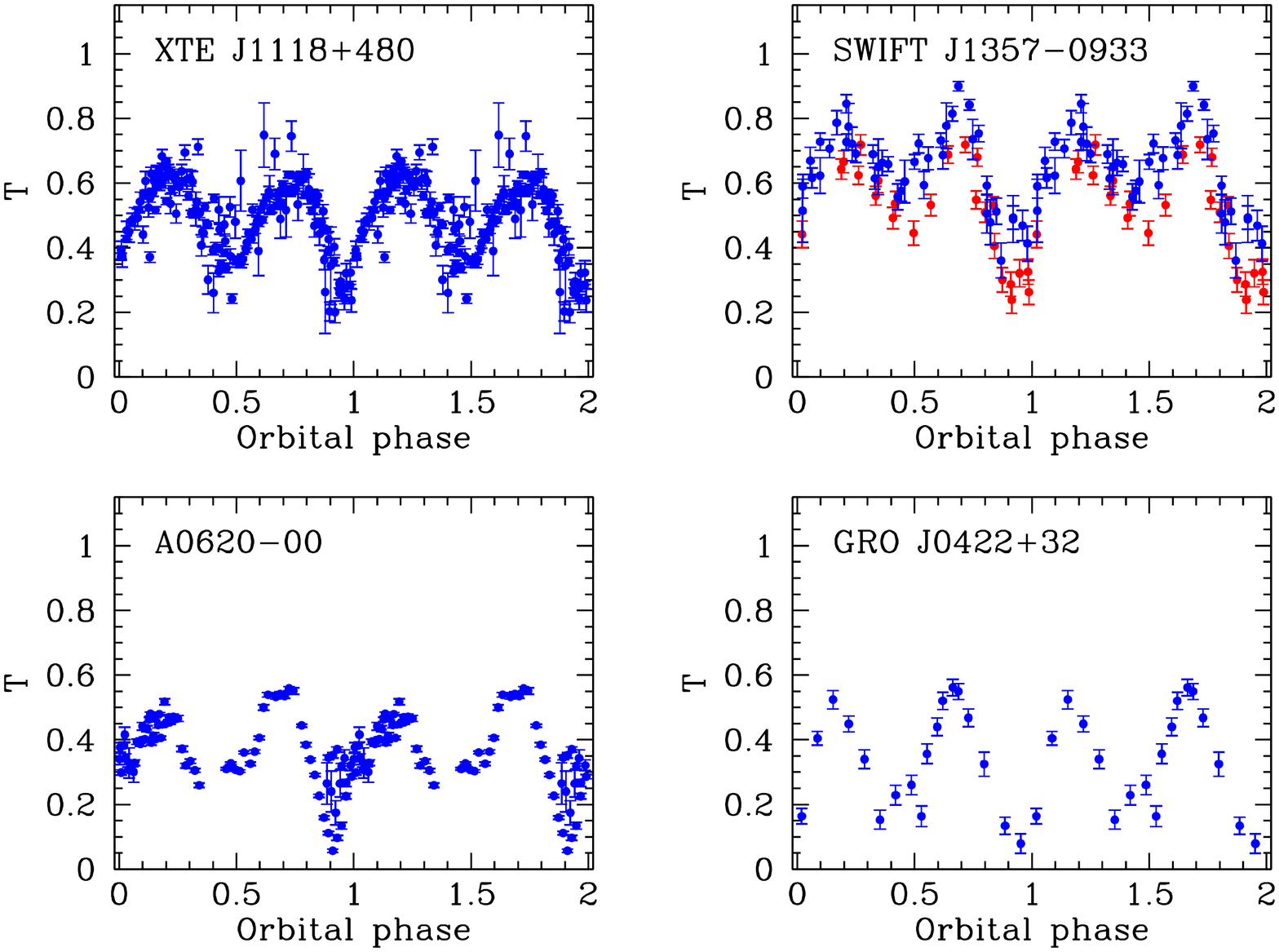}
    \caption{Orbital modulation of $T$ in a sample of BH XRTs. Red and blue points in the upper right panel represents two different data sets: 2013 VLT data from Torres et al. (2015) and 2014 GTC data from Mata S\'anchez et al. (2015), respectively. 
 The J1118 and A0620 data have been folded using ephemeris from Gonz\'alez Hern\'andez et al. (2014), while 
 the J0422 data using those from Webb et al. (2000). In the case of J1357 we have used  our new period 
 determination $P_{\rm orb}=0.106969$ d and a tentative time of inferior conjunction of the companion star of HJD 2456396.6617, 
 obtained by imposing that the deepest minimum is centered at phase 0.95. } 
    \label{fig:a3}
\end{figure}

All systems display a clear double-hump modulation of $T$ versus orbital phase, 
with maxima at phases 0.2 and 0.7 and minima at 0.45 and 0.95. The phasing suggests that the minima are caused by 
the crossing of S-waves (i.e. narrow emission-line components associated to the hot spot and the companion star)  
through the center of the line profile and this is confirmed by trailed spectrogram plots (see e.g. \citealt{marsh94,zurita16}). 
A sine fit with the period fixed to $0.5\times P_{\rm orb}$ yields a characteristic amplitude $\Delta T=\pm0.15$. Given the double-sine 
shape of the modulation,  $T$ values obtained from average spectra with $\gtrsim$ 50 per cent phase coverage must be close to 
the orbital mean and, therefore, not significantly biased. 

On the other hand, we observe that the phase 0.95 minimum appears deeper than the minimum at phase 0.45. This can be 
explained by opacity effects since the hot-spot becomes obscured by the outer disc rim around phase 0.4 thus making the 
filling-in of the double peak trough less pronounced. As a matter of fact, the different minima can be exploited to determine the orbital 
period in systems where the companion star is totally veiled by the accretion disc. For example, we have performed a period analysis 
of the time evolution of $T$ in J1357. Fig.~\ref{fig:a3} presents the resulting Lomb-Scargle periodogram. 
The frequency of the highest peak corresponds to a period of 0.005348455$\pm$0.0000115 d, where the error is taken 
as the sigma of a Gaussian fit to the peak. Because of the double-humped shape of the $T$ curve we adopt twice 
this value as the true orbital period i.e. $P_{\rm orb}=0.106969\pm0.000023$ d. 
This is consistent but much more precise than previous reports in \cite{corral13} and \cite{mata15} because of the 
14 months elapsed between the VLT and GTC  observations. 
It should be noted that only the peak with the highest power produces a phase folded $T$ curve with unequal minima 
(see Fig.~\ref{fig:a2}). This confirms that the next high peaks are caused by aliasing and can be discarded. 

\begin{figure}
	\includegraphics[angle=-90,width=\columnwidth]{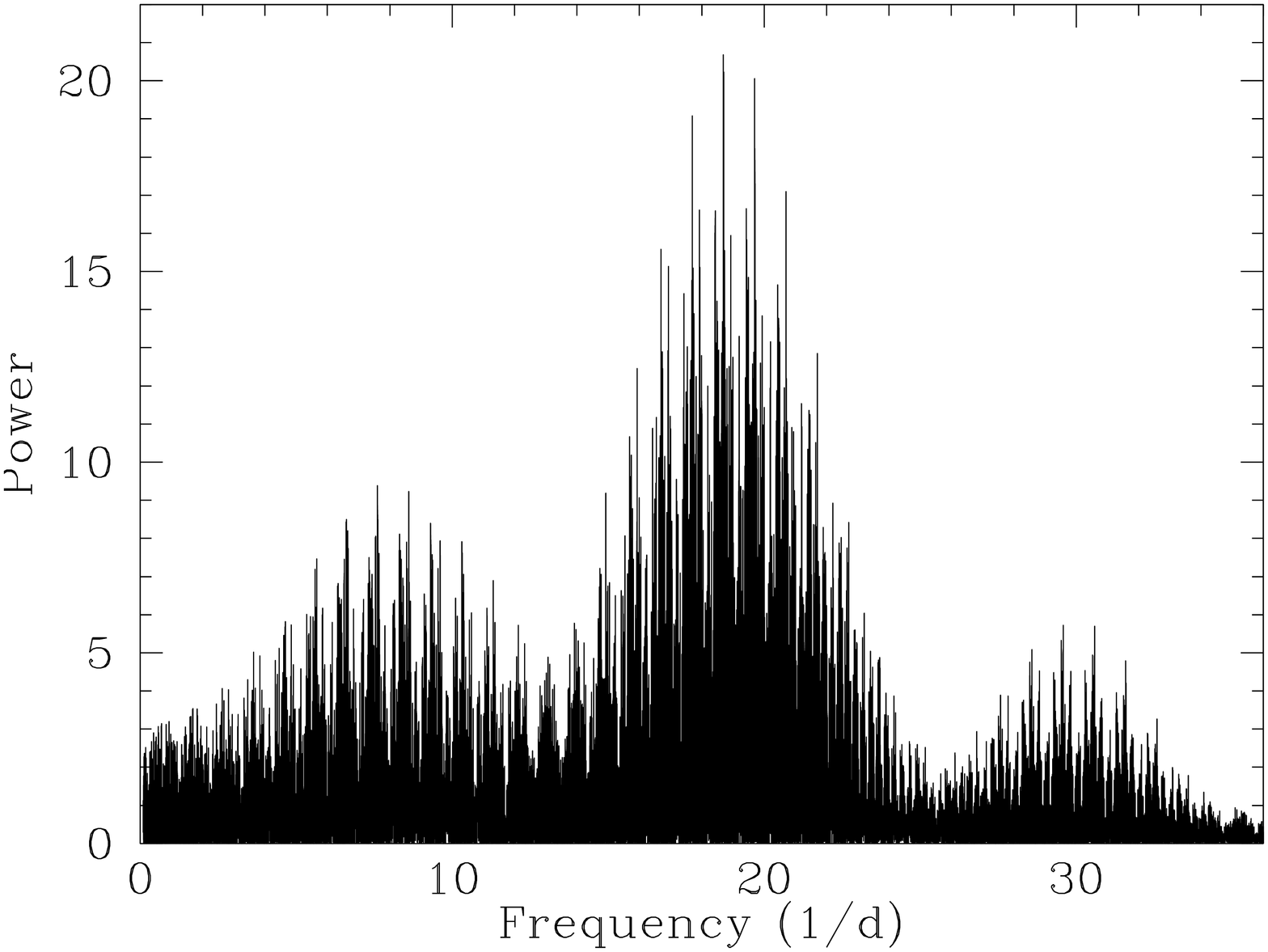}
    \caption{Lomb-Scargle periodogram of $T$ measurements from the 2013 VLT and 2014 GTC campaigns of J1357. 
    The highest peak corresponds to our selected  0.106969 d period.}
    \label{fig:a4}
\end{figure}

\section{Inclination reports disregarded from the calibration sample}
\label{ap:dismissed_inclination}

In the particular case of A0620 we have dismissed the works of \cite{shahbaz94} 
($i=31-54^{\circ}$) and \cite{gelino01a} ($i=41\pm3^{\circ}$) because disc contamination is 
neglected (although the latter attempt to account for light curve asymmetries using a dark spot 
stellar model). \cite{froning01}, on the other hand, do not provide a  
useful constrain to the disc contribution on their H-band light curves since the 
very wide range quoted ($i=38-75^{\circ}$) is virtually uninformative. Finally, we regard the 
result of \cite{haswell93} ($i=62-74^{\circ}$) as dubious as it relies on the 
identification of a tentative grazing eclipse that was never confirmed in 
subsequent higher quality light curves. We believe that a transient sharp asymmetry, 
associated for example to a running superhump wave (see an example in \citealt{zurita02}) 
provides a more plausible explanation to the data. Note that, 
despite Haswell's data expand over several nights the authors applied a weighted 
process to produce their phased light curves because "the quality of the data varied 
significantly from night to night" (sic) so it seems possible that a sharp asymmetry 
on a high-quality night may stand up in the mean light curve.  
 
Regarding N Mus 91, we have dismissed the work of \cite{shahbaz97} because the 
accretion disc contribution is neglected. Besides, the limited data quality results 
in a poorly constrained inclination value $i=39-64^{\circ}$. 
Likewise, we dismiss \cite{orosz96} ($i=54-65^{\circ}$) and \cite{gelino01b}  
($i=54\pm2^{\circ}$) because the contamination by non-stellar sources is crudely 
modeled or neglected. These results are superseded by a later paper of 
the same group \citep{wu16}, that also incorporates a more 
extended photometric database from subsequent epochs. Wu et al. apply a careful filtering, 
selecting only light curves during passive state. In particular, a re-analisis of  
the light curves of \cite{orosz96} and \cite{gelino01b}  is performed, after correcting for 
 accretion disc contamination and including a hot-spot contribution. 
They consistently find $i=43^{\circ}$ in all cases. In view of all this we decide to 
choose the result of \cite{wu16} as the best estimate available for the inclination in 
N Mus 91.  

In the case of GS2000 we do not consider the work of \cite{beekman96} 
 ($i=43-69^{\circ}$) because the accretion disc contribution is neglected when fitting 
ellipsoidal models to a limited quality H-band light curve. 

As for J1118 we systematically ignore inclination measurements obtained 
when the system was in the decay phase. These are: \cite{mcclintock01} ($i\geq55^{\circ}$), 
\cite{wagner01} ($i=81\pm2^{\circ}$) and \cite{zurita02} ($i=71-82^{\circ}$). 

Regarding J1550 we did not consider the work of \cite{orosz02} ($i=67-74^{\circ}$) 
because it is superseded by our selected paper \cite{orosz11}  ($i=74.7\pm3.8^{\circ}$). 
The latter work presents fits to new higher quality NIR light curves, 
together with a re-analysis of the older light curves from \cite{orosz02}, including 
accretion disc contamination.   

Finally, in the case of Cen X-4, we have dismissed \cite{shahbaz93}
($i=31-54^{\circ}$) because the accretion disc contribution is ignored. This work is superseded 
by \cite{khargharia10}, one of our selected results, where the 
accretion disc contribution is estimated by means of NIR spectroscopy. 


\bsp	
\label{lastpage}
\end{document}